\documentclass[english]{article}
\usepackage[T1]{fontenc}
\usepackage[latin9]{inputenc}
\usepackage{array}
\usepackage{units}
\usepackage{multirow}
\usepackage{amsmath}
\usepackage{amssymb}
\usepackage{stmaryrd}
\usepackage{graphicx}
\usepackage[numbers]{natbib}

\makeatletter

\newcommand*\LyXZeroWidthSpace{\hspace{0pt}}
\providecommand{\tabularnewline}{\\}

\newcommand{\lyxaddress}[1]{
	\par {\raggedright #1
	\vspace{1.4em}
	\noindent\par}
}
\newenvironment{lyxlist}[1]
	{\begin{list}{}
		{\settowidth{\labelwidth}{#1}
		 \setlength{\leftmargin}{\labelwidth}
		 \addtolength{\leftmargin}{\labelsep}
		 }}
	{\end{list}}

\@ifundefined{date}{}{\date{}}
\usepackage{placeins}
\usepackage{array}
\usepackage{nicefrac}
\usepackage{adjustbox}
\usepackage{floatrow}
\usepackage[font=small]{caption}
\newcommand{\deq}{\overset{d}{=}}
\newcommand{\Prob}{\mathbb{P}}

\makeatother

\usepackage{babel}
\begin{document}
\title{Degrees, Levels, and Profiles of Contextuality}
\author{Ehtibar N. Dzhafarov$^{1}$ \& Víctor H. Cervantes$^{2}$}
\maketitle

\lyxaddress{\begin{center}
\textsuperscript{1}Purdue University, USA, ehtibar@purdue.edu\\
\textsuperscript{2}University of Illinois at Urbana-Champaign, USA,
victorhc@illinois.edu
\par\end{center}}
\begin{abstract}
We introduce a new notion, that of a \emph{contextuality profile}
of a system of random variables. Rather than characterizing a system's
contextuality by a single number, its overall \emph{degree of contextuality},
we show how it can be characterized by a curve relating degree of
contextuality (including nonlocality, as a special case) to \emph{level}
at which the system is considered,
\[
\begin{array}{c|c|c|c|c|c|c|c}
\textnormal{level} & 1 & \cdots & n-1 & n>1 & n+1 & \cdots & N\\
\hline \textnormal{degree} & 0 & \cdots & 0 & d_{n}>0 & d_{n+1}\geq d_{n} & \cdots & d_{N}\geq d_{N-1}
\end{array},
\]
where $N$ is the maximum number of variables per system's context.
A system is represented at level $n$ if one only considers the joint
distributions with $k\leq n$ variables, ignoring higher-order joint
distributions. We show that the level-wise contextuality analysis
can be used in conjunction with any well-constructed measure of contextuality.
We present a method of concatenated systems to explore contextuality
profiles systematically, and we apply it to the contextuality profiles
for three major measures of contextuality proposed in the literature.

\end{abstract}
There is a consensus that it is not sufficiently informative to merely
ascertain whether a system of random variables is contextual. It is
also desirable and useful to measure its \emph{degree of contextuality}
(for a recent survey of the contextuality literature, see Ref. \citep{BCGKL}).
We recently proposed that it is also of interest to determine a system's
\emph{first level of contextuality}. The meaning of this term is as
follows. Let the joint distribution of any $k$ variables in the system
be called a $k$-\emph{marginal}. For any $n$, if one characterizes
the system by all its $k$-marginals with $k\leq n$, ignoring the
higher-level marginals, we say that the system is \emph{represented
at level} $n$. A contextual system is always noncontextual at level
1, and there is a lowest level $n>1$ at which it becomes contextual.
This $n$ is the system's \emph{first level of contextuality}, and
at this level the degree of contextuality, $d_{n}$, can be measured
in several known ways. In Ref. \citep{book2026}, this procedure is
described in detail for the measure based on the $L_{1}$-distance
between a point representing a system at a given level and the corresponding
noncontextuality polytope. 

We now show that the level-wise analysis can be used in conjunction
with any other measure of contextuality (notably, the contextual fraction,
and the measure involving ``negative probabilities'', both extended
to apply to systems with disturbance). Moreover, for a system determined
to be contextual at level $n$, one can continue to measure its contextuality
degree at levels $n+1,n+2,\ldots,N$ (where $N$ is the maximum number
of variables per system's context). In this way one characterizes
a system by a \emph{vector of contextuality values} 
\begin{equation}
\begin{array}{c|c|c|c|c|c|c|c}
\textnormal{level} & 1 & \cdots & n-1 & n & n+1 & \cdots & N\\
\hline \textnormal{degree} & 0 & \cdots & 0 & d_{n} & d_{n+1} & \cdots & d_{N}
\end{array}
\end{equation}
that can be called the system's \emph{contextuality profile}. We discuss
ways of exploring the patterns of contextuality systematically. For
well-constructed measures of contextuality, $d_{k}$ is nondecreasing
in $k.$ This is true for the three measures mentioned above. The
tendency of $d_{k}$ to increase with $k$ is minimal for the ``negative
probabilities'' measure and maximal for the $L^{1}$ distance measure,
with the contextual fraction falling in between.

The structure of the paper is as follows. In Section \ref{sec:Definition-of-contextuality},
we remind the reader of the definition of contextuality. Section \ref{sec:Level-wise-representations}
introduces the notion of a level-$n$ representation of a system of
random variables. Section \ref{sec:Profiles-of-contextuality} presents
the main idea of the paper: given a measure of contextuality, to define
a system's contextuality profile as the sequence of its contextuality
values at different levels. In Section \ref{sec:Three-measures-of},
we discuss three specific measures of contextuality that have been
proposed in the contextuality literature. In Section \ref{sec:Concatenated-systems},
we propose a method of concatenated systems for exploring how fast
contextuality profiles grow from one level to another. Section \ref{sec:Contextuality-profiles-for}
presents the results of applying this method to the three measures
of contextuality just mentioned. Section \ref{sec:Outlines} contains
proofs of the regularities discovered in Section \ref{sec:Contextuality-profiles-for}.
In Section \ref{sec:Other-systems}, we show that none of the three
measures of contextuality is a function of another, by studying contextuality
profiles of a special class of systems. Section \ref{sec:Conclusion}
offers a summary and some questions for future work.

\section{\label{sec:Definition-of-contextuality}Definition of contextuality}

Consider a generic example of a system:
\begin{equation}
\begin{array}{|c|c|c|c|c||c|}
\hline  & R_{2}^{1} & R_{3}^{1} &  &  & c=1\\
\hline R_{1}^{2} & R_{2}^{2} &  & R_{4}^{2} & R_{5}^{2} & 2\\
\hline R_{1}^{3} & R_{2}^{3} & R_{3}^{3} & R_{4}^{3} &  & 3\\
\hline R_{1}^{4} & R_{2}^{4} &  &  & R_{5}^{4} & 4\\
\hline\hline q=1 & 2 & 3 & 4 & 5 & \mathcal{R}
\\\hline \end{array}\:.\label{eq:example system}
\end{equation}
We use our usual notation here. $R_{q}^{c}$ is a random variable
recorded in context $c$ and answering question $q$. If $R_{q}^{c}$
is defined for a given $\left(q,c\right)$, i.e., if the cell is not
empty, we write $q\Yleft c$ (question $q$ is answered in context
$c$). The main property of a system of random variables is that all
the variables in each row (sharing a context) are\emph{ jointly distributed},
whereas no two variables from different rows are jointly distributed
(they are \emph{stochastically unrelated}).

A system is \emph{consistently connected} if, for any $q\Yleft c,c'$,
\begin{equation}
R_{q}^{c}\overset{d}{=}R_{q}^{c'},
\end{equation}
that is, any two variables answering the same question are identically
distributed. A system is \emph{undisturbed} (or \emph{strongly consistently
connected}) if, for any subset of questions $q_{1},\ldots,q_{k}\Yleft c,c'$,
\begin{equation}
\left(R_{q_{1}}^{c},\ldots,R_{q_{k}}^{c}\right)\overset{d}{=}\left(R_{q_{1}}^{c'},\ldots,R_{q_{k}}^{c'}\right).
\end{equation}
For instance, if the system $\mathcal{R}$ above is undisturbed, then
\begin{equation}
\left(R_{1}^{2},R_{2}^{2},R_{4}^{2}\right)\overset{d}{=}\left(R_{1}^{3},R_{2}^{3},R_{4}^{3}\right),\left(R_{2}^{1},R_{3}^{1}\right)\overset{d}{=}\left(R_{2}^{3},R_{3}^{3}\right),\text{etc}.
\end{equation}

The notion of contextuality, extended to include disturbed systems,
is described in detail in our previous publications (e.g., Refs. \citep{book2026,KujDzhMeasures}).
Put briefly, and using the system $\mathcal{R}$ in (\ref{eq:example system}),
we attempt to construct its \emph{probabilistic coupling} 
\begin{equation}
\begin{array}{|c|c|c|c|c||c|}
\hline  & S_{2}^{1} & S_{3}^{1} &  &  & c=1\\
\hline S_{1}^{2} & S_{2}^{2} &  & S_{4}^{2} & S_{5}^{2} & 2\\
\hline S_{1}^{3} & S_{2}^{3} & S_{3}^{3} & S_{4}^{3} &  & 3\\
\hline S_{1}^{4} & S_{2}^{4} &  &  & S_{5}^{4} & 4\\
\hline\hline q=1 & 2 & 3 & 4 & 5 & S
\\\hline \end{array}\:,\label{eq:coupling exmaple}
\end{equation}
in which all variables are jointly distributed (not just within contexts
but overall), subject to the following constraints: 
\begin{lyxlist}{00.00.0000}
\item [{(a)}] The variables in the rows of $S$ are jointly distributed
as the variables in the corresponding row of $\mathcal{R}$, e.g.,
\[
\left(S_{2}^{1},S_{3}^{1}\right)\deq\left(R_{2}^{1},R_{3}^{1}\right),\left(S_{1}^{2},S_{2}^{2},S_{4}^{2},S_{5}^{2}\right)\deq\left(R_{1}^{2},R_{2}^{2},R_{4}^{2},R_{5}^{2}\right),\text{etc}.
\]
\item [{(b)}] Any two variables answering the same question (e.g., $S_{2}^{1}$
and $S_{2}^{3}$) coincide with the maximal possible probability.
\end{lyxlist}
If a coupling of $\mathcal{R}$ satisfying (a) and (b) exists, the
system $\mathcal{R}$ is \emph{noncontextual}, otherwise it is \emph{contextual}. 

For consistently connected systems, where the variables answering
the same question are identically distributed, the maximal probability
of coinciding, e.g., of $\left[S_{2}^{1}=S_{2}^{3}\right]$, equals
1. That is, the requirement (b) then simply makes all variables answering
the same question identical. In this case one can replace the couplings
of $\mathcal{R}$ with \emph{reduced couplings},
\begin{equation}
\left(S_{1},S_{2},S_{3},S_{4},S_{5}\right).
\end{equation}
The system $\mathcal{R}$ then is noncontextual if and only if there
exists a reduced coupling such that 
\[
\left(S_{2},S_{3}\right)\deq\left(R_{2}^{1},R_{3}^{1}\right),\left(S_{1},S_{2},S_{4},S_{5}\right)\deq\left(R_{1}^{2},R_{2}^{2},R_{4}^{2},R_{5}^{2}\right),\text{etc}.
\]

Let $\text{\textbf{v}}$ be a vector of probabilities (all vectors
in this paper are columns unless shown as transposed, $\mathbf{\mathbf{v}^{\intercal}}$)
that represents the constraints imposed on the distribution of the
variables in the system's coupling. For now, it is not important precisely
how it is constructed. Suffice it to stipulate that $\mathbf{v}$
uniquely determines the joint distributions of the variables in each
context, and also ensures that all same-question pairs of variables
coincide with maximal probabilities. The existence of the coupling
with properties (a)-(b) then means that there is a vector $\mathbf{x}$
satisfying the following matrix equation:
\begin{equation}
\mathbf{M}\,\mathbf{x}=\mathbf{v},\quad\mathbf{x}\geq\mathbf{0}\quad\left(\mathbf{1}^{\intercal}\mathbf{x}=1\right).\label{eq:noncontextuality equation}
\end{equation}
Here, $\mathbf{x}\geq\mathbf{0}$ means that every component of $\mathbf{x}$
is nonnegative; $\mathbf{1}$ is a vector of $1$s (so that $\mathbf{1}^{\intercal}\mathbf{x}$
is the sum of the components of $\mathbf{x}$). The components of
$\mathbf{x}$ are probabilities of all possible values of the coupling
$S$. Thus, in our example (\ref{eq:coupling exmaple}), if all variables
are dichotomous, there are $2^{13}$ possible values, such as
\[
\left[S_{2}^{1}=1,S_{3}^{1}=0,S_{1}^{2}=0,\ldots,S_{5}^{4}=1\right],
\]
and $\mathbf{x}$ contains the same number of probabilities. $\mathbf{M}$
is a Boolean incidence matrix which tells us which components of $\mathbf{x}$
sum to a given probability in $\mathbf{v}$. The reason $\mathbf{1}^{\intercal}\mathbf{x}=1$
is shown in (\ref{eq:noncontextuality equation}) parenthetically
is that this condition is satisfied automatically due to the fact
that $\mathbf{v}$ represents probability distributions.

\section{\label{sec:Level-wise-representations}Level-wise representations}

Let context $c$ of a system $\mathcal{R}$ contain $N$ random variables
\[
\left(R_{q_{1}}^{c},\ldots,R_{q_{N}}^{c}\right)=R^{c}.
\]
The representation of this context at level $n\leq N$ is defined
as a system $\mathcal{R}^{c}\left[n\right]$ with $\binom{N}{n}$
contexts each containing a distinct $n$-tuple of variables selected
from $R^{c}$. For $n>N$, the level $n$ representation $\mathcal{R}^{c}\left[n\right]$
consists of a single context containing $R^{c}$. (The reason we denote
the row of variables $R^{c}$, in italics, but denote the representing
system $\mathcal{R}^{c}\left[n\right]$, in script letters, is that
we use italics for sets of variables when they are jointly distributed
and script letters when they are not, or are not necessarily. We use
this notation convention throughout this paper.)

The representation of a system $\mathcal{R}$ at level $n$ is a system
$\mathcal{R}\left[n\right]$ consisting of the representation of all
contexts of $\mathcal{R}$ at level $n$.

As an example, let us consider the second context of system $\mathcal{R}$
in (\ref{eq:example system}):
\begin{equation}
\begin{array}{|c|c|c|c|c||c|}
\hline R_{1}^{2} & R_{2}^{2} &  & R_{4}^{2} & R_{5}^{2} & c=2\\
\hline\hline q=1 & 2 & 3 & 4 & 5 & R^{2}
\\\hline \end{array}\:.
\end{equation}
Skipping level 1, which will be discussed later, the representation
of this row at level 2 is
\begin{equation}
\begin{array}{|c|c|c|c|c||c|}
\hline\hline R_{1}^{2.1} & R_{2}^{2.1} &  &  &  & 2.1\\
\hline R_{1}^{2.2} &  &  & R_{4}^{2.2} &  & 2.2\\
\hline R_{1}^{2.3} &  &  &  & R_{5}^{2.3} & 2.3\\
\hline  & R_{2}^{2.4} &  & R_{4}^{2.4} &  & 2.4\\
\hline  & R_{2}^{2.5} &  &  & R_{5}^{2.5} & 2.5\\
\hline  &  &  & R_{4}^{2.6} & R_{5}^{2.6} & 2.6\\
\hline\hline q=1 & 2 & 3 & 4 & 5 & \mathcal{R}^{2}\left[2\right]
\\\hline \end{array}\:,
\end{equation}
where, for any $q,q'\Yleft c=2$,
\begin{equation}
\left(R_{q}^{2.x},R_{q'}^{2.x}\right)\overset{d}{=}\left(R_{q}^{2},R_{q'}^{2}\right).
\end{equation}
In other words, each of the $\binom{4}{2}=6$ rows of the new matrix
is obtained by picking from the row $R^{2}$ a subset of two variables,
and creating their distributional copy. Clearly, the system $\mathcal{R}^{2}\left[2\right]$
representing the row $R^{2}$ at level 2 is an undisturbed system:
e.g.,
\begin{equation}
R_{2}^{2.1}\overset{d}{=}R_{2}^{2.4}\overset{d}{=}R_{2}^{2.5}\overset{d}{=}R_{2}^{2},R_{4}^{2.2}\overset{d}{=}R_{4}^{2.4}\overset{d}{=}R_{4}^{2.6}\overset{d}{=}R_{4}^{2},\text{etc}.
\end{equation}
Moreover, the system $\mathcal{R}^{2}\left[2\right]$ is noncontextual
because it has a coupling coinciding in distribution with $R^{2}$.
The system $\mathcal{R}^{2}\left[2\right]$ has the same individual
and pairwise distributions as the row $R^{2}$, but $\mathcal{R}^{2}\left[2\right]$
contains no higher-order distributions (no triples, quadruples, etc.)
because joint distributions only exist within but not across the contexts.

The representation $\mathcal{R}^{2}\left[3\right]$ of the row $R^{2}$
at level 3 is obtained analogously, by picking from this row $\binom{4}{3}=4$
possible triples of variables and creating their distributional copies:
\begin{equation}
\begin{array}{|c|c|c|c|c||c|}
\hline R_{1}^{2.1} & R_{2}^{2.1} &  & R_{4}^{2.1} &  & 2.1\\
\hline R_{1}^{2.2} & R_{2}^{2.2} &  &  & R_{5}^{2.2} & 2.2\\
\hline R_{1}^{2.3} &  &  & R_{4}^{2.3} & R_{5}^{2.3} & 2.3\\
\hline  & R_{2}^{2.4} &  & R_{4}^{2.4} & R_{5}^{2.4} & 2.4\\
\hline\hline q=1 & 2 & 3 & 4 & 5 & \mathcal{R}^{2}\left[3\right]
\\\hline \end{array}\:.
\end{equation}
This system, too, is undisturbed by construction, and it is noncontextual
as it has a coupling that coincides in distribution with $R^{2}$. 

Finally, the level 4 representation of the row $R^{2}$ simply coincides
with it, because there is only one quadruple we can select from $R^{2}$:
\begin{equation}
\begin{array}{l}
\begin{array}{|c|c|c|c|c||c|}
\hline R_{1}^{2} & R_{2}^{2} &  & R_{4}^{2} & R_{5}^{2} & c=2\\
\hline\hline q=1 & 2 & 3 & 4 & 5 & \mathcal{R}^{2}\left[4\right]=R^{2}\left[4\right]
\\\hline \end{array}\\
\\=\begin{array}{|c|c|c|c|c||c|}
\hline R_{1}^{2} & R_{2}^{2} &  & R_{4}^{2} & R_{5}^{2} & c=2\\
\hline\hline q=1 & 2 & 3 & 4 & 5 & R^{2}
\\\hline \end{array}\:.
\end{array}
\end{equation}
This system is trivially undisturbed and trivially noncontextual.

The higher-level representations $\mathcal{R}^{2}\left[5\right]$,
$\mathcal{R}^{2}\left[6\right]$, etc. of $R^{2}$ also simply coincide
with $R^{2}$, which is explained as follows. A representation at
level $n>1$ should always be taken as cumulative, to include not
only $n$-tuples but also all lower-level tuples. However, if $n$-tuples
exist (the original row contains no less than $n$ variables), inclusion
or exclusion of the lower-level tuples never influences the contextuality
status of the representing system (i.e., whether the system is contextual
or noncontextual) or any reasonable measure of the degree of contextuality
if it is contextual. So, e.g., the system $\mathcal{R}^{2}\left[3\right]$
with an added row $c=2.5$ that contains the pair $\left(R_{1}^{2.2},R_{2}^{2.2}\right)$,
\begin{equation}
\begin{array}{|c|c|c|c|c||c|}
\hline R_{1}^{2.1} & R_{2}^{2.1} &  & R_{4}^{2.1} &  & 2.1\\
\hline R_{1}^{2.2} & R_{2}^{2.2} &  &  & R_{5}^{2.2} & 2.2\\
\hline R_{1}^{2.3} &  &  & R_{4}^{2.3} & R_{5}^{2.3} & 2.3\\
\hline  & R_{2}^{2.4} &  & R_{4}^{2.4} & R_{5}^{2.4} & 2.4\\
\hline R_{1}^{2.2} & R_{2}^{2.2} &  &  &  & 2.5\\
\hline\hline q=1 & 2 & 3 & 4 & 5 & \mathcal{R}^{2+}\left[3\right]
\\\hline \end{array}\:,
\end{equation}
is equivalent to $\mathcal{R}^{2}\left[3\right]$ in any considerations
of contextuality. This is a general property of undisturbed systems.
That is why we defined $\mathcal{R}^{2}\left[3\right]$ as containing
only triples rather than also pairs and singles, and $\mathcal{R}^{2}\left[4\right]$
as containing only quadruples, ignoring triples, pairs, and singles.
However, on the next level, $5$, no quintuples of variables exist,
so we have to include the highest existing tuple, which in this case
is the quadruple:
\begin{equation}
\begin{array}{l}
\begin{array}{|c|c|c|c|c||c|}
\hline R_{1}^{2} & R_{2}^{2} &  & R_{4}^{2} & R_{5}^{2} & c=2\\
\hline\hline q=1 & 2 & 3 & 4 & 5 & \mathcal{R}^{2}\left[5\right]=R^{2}\left[5\right]
\\\hline \end{array}\\
\\=\begin{array}{|c|c|c|c|c||c|}
\hline R_{1}^{2} & R_{2}^{2} &  & R_{4}^{2} & R_{5}^{2} & c=2\\
\hline\hline q=1 & 2 & 3 & 4 & 5 & R^{2}
\\\hline \end{array}\:.
\end{array}
\end{equation}
By the same logic, the representation of the first row of the system
$\mathcal{R}$,
\begin{equation}
\begin{array}{|c|c|c|c|c||c|}
\hline  & R_{2}^{1} & R_{3}^{1} &  &  & c=1\\
\hline\hline q=1 & 2 & 3 & 4 & 5 & R^{1}
\\\hline \end{array}\:,
\end{equation}
is one and the same at all levels $n>1$:
\begin{equation}
R^{1}=\mathcal{R}^{1}\left[2\right]=\mathcal{R}^{1}\left[3\right]=\cdots
\end{equation}

Let us now explain why we do not consider level 1 representations.
In fact, we do include them in the above-given definition of the level-wise
representations, but they are always trivially noncontextual, requiring
no separate analysis. For our example, the row $R^{2}$, the level
1 representation is
\begin{equation}
\begin{array}{|c|c|c|c|c||c|}
\hline R_{1}^{2.1} &  &  &  &  & 2.1\\
\hline  & R_{2}^{2.2} &  &  &  & 2.2\\
\hline  &  &  & R_{4}^{2.3} &  & 2.3\\
\hline  &  &  &  & R_{5}^{2.4} & 2.4\\
\hline\hline q=1 & 2 & 3 & 4 & 5 & \mathcal{R}^{2}\left[1\right]
\\\hline \end{array}\:.
\end{equation}
A row with only one variable in it can be removed from any system
without affecting its contextuality status. 

With the algorithm specified, the level 4 representation of the entire
system $\mathcal{R}$ is
\[
\begin{array}{|c|c|c|c|c||c|}
\hline  & R_{2}^{1} & R_{3}^{1} &  &  & c=1\\
\hline R_{1}^{2} & R_{2}^{2} &  & R_{4}^{2} & R_{5}^{2} & 2\\
\hline R_{1}^{3} & R_{2}^{3} & R_{3}^{3} & R_{4}^{3} &  & 3\\
\hline R_{1}^{4} & R_{2}^{4} &  &  & R_{5}^{4} & 4\\
\hline\hline q=1 & 2 & 3 & 4 & 5 & \mathcal{R}\left[4\right]=\mathcal{R}
\\\hline \end{array}\:,
\]
and its level 3 and level 2 representations are, respectively,
\begin{equation}
\begin{array}{|c|c|c|c|c||c|}
\hline  & R_{2}^{1} & R_{3}^{1} &  &  & c=1\\
\hline\hline R_{1}^{2.1} & R_{2}^{2.1} &  & R_{4}^{2.1} &  & 2.1\\
\hline R_{1}^{2.2} & R_{2}^{2.2} &  &  & R_{5}^{2.2} & 2.2\\
\hline R_{1}^{2.3} &  &  & R_{4}^{2.3} & R_{5}^{2.3} & 2.3\\
\hline  & R_{2}^{2.4} &  & R_{4}^{2.4} & R_{5}^{2.4} & 2.4\\
\hline\hline R_{1}^{3.1} & R_{2}^{3.1} & R_{3}^{3.1} &  &  & 3.1\\
\hline R_{1}^{3.2} & R_{2}^{3.2} &  & R_{4}^{3.2} &  & 3.2\\
\hline R_{1}^{3.3} &  & R_{3}^{3.3} & R_{4}^{3.3} &  & 3.3\\
\hline  & R_{2}^{3.4} & R_{3}^{3.4} & R_{4}^{3.4} &  & 3.4\\
\hline\hline R_{1}^{4} & R_{2}^{4} &  &  & R_{5}^{4} & 4\\
\hline\hline q=1 & 2 & 3 & 4 & 5 & \mathcal{R}\left[3\right]
\\\hline \end{array}\:.
\end{equation}
and

\begin{equation}
\begin{array}{|c|c|c|c|c||c|}
\hline  & R_{2}^{1} & R_{3}^{1} &  &  & c=1\\
\hline\hline R_{1}^{2.1} & R_{2}^{2.1} &  &  &  & 2.1\\
\hline R_{1}^{2.2} &  &  & R_{4}^{2.2} &  & 2.2\\
\hline R_{1}^{2.3} &  &  &  & R_{5}^{2.3} & 2.3\\
\hline  & R_{2}^{2.4} &  & R_{4}^{2.4} &  & 2.4\\
\hline  & R_{2}^{2.5} &  &  & R_{5}^{2.5} & 2.5\\
\hline  &  &  & R_{4}^{2.6} & R_{5}^{2.6} & 2.6\\
\hline\hline R_{1}^{3.1} & R_{2}^{3.1} &  &  &  & 3.1\\
\hline R_{1}^{3.2} &  & R_{3}^{3.2} &  &  & 3.2\\
\hline R_{1}^{3.3} &  &  & R_{4}^{3.3} &  & 3.3\\
\hline  & R_{2}^{3.4} & R_{3}^{3.4} &  &  & 3.4\\
\hline  & R_{2}^{3.5} &  & R_{4}^{3.5} &  & 3.5\\
\hline  &  & R_{3}^{3.6} & R_{4}^{3.6} &  & 3.6\\
\hline\hline R_{1}^{4.1} & R_{2}^{4.1} &  &  &  & 4.1\\
\hline R_{1}^{4.2} &  &  &  & R_{5}^{4.2} & 4.2\\
\hline  & R_{2}^{4.3} &  &  & R_{5}^{4.3} & 4.3\\
\hline\hline q=1 & 2 & 3 & 4 & 5 & \mathcal{R}\left[2\right]
\\\hline \end{array}\:.
\end{equation}

\section{\label{sec:Profiles-of-contextuality}Profiles of contextuality}

Let us assume that we have a measure of contextuality that applies
to any system $\mathcal{R}$ (from a sufficiently broad class of systems).
Let us denote its value by $\mathsf{deg}\mathcal{R}$. The main idea
of this paper is this: contextuality values of the level-wise representations
of $\mathcal{R}$, 
\begin{equation}
\mathsf{deg}\mathcal{R}\left[1\right]=0,\mathsf{deg}\mathcal{R}\left[2\right]=d_{2},\ldots,\mathsf{deg}\mathcal{R}\left[N\right]=d_{N},
\end{equation}
can be considered the \emph{contextuality profile} of the system $\mathcal{R}$.
Here, $N$ is the maximal number of variables in a row of $\mathcal{R}$.
The values of $d_{N+1}$, $d_{N+2}$, etc., need not be considered
because they always equal $d_{N}$. We include the uninformative $\mathsf{deg}\mathcal{R}\left[1\right]=0$
as the ``anchoring point'' of a profile, primarily for aesthetic
reasons.

\begin{figure}
\begin{centering}
\includegraphics[scale=0.65]{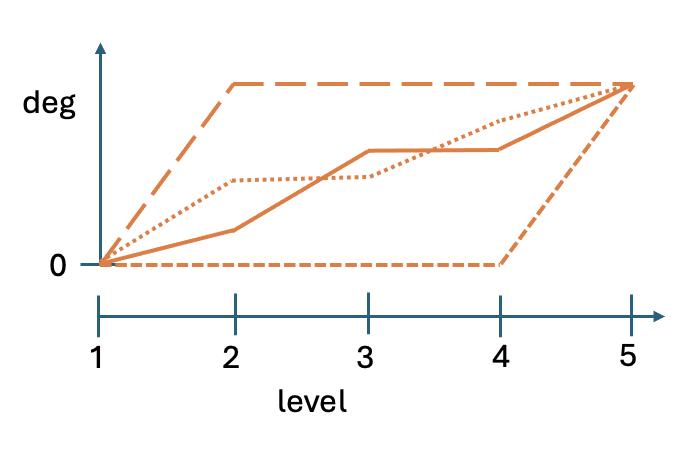}
\par\end{centering}
\caption{\label{fig:Four-possible-contextuality}Four possible contextuality
profiles with the same final degree of contextuality at level 5.}
\end{figure}

Figure \ref{fig:Four-possible-contextuality} presents hypothetical
contextuality profiles for four systems with $N=5$. Observe that
at level 5 all four profiles have the same value. This common value
is the contextuality degree that our measure $\mathsf{deg}$ will
show for all four systems, because level 5 representations of these
systems coincide with the systems themselves. The existing ways of
contextuality analysis therefore would treat these four systems as
essentially indistinguishable.

In Ref. \citep{book2026}, this idea is partially implemented for
the measure of contextuality that we called ``hierarchical.'' The
implementation involves levels of consideration, but the process described
there stops at the first contextual level (the smallest $n$ with
$d_{n}>0$). In essence, for a contextual system, this merely replaces
a point measure of contextuality (the single number $d_{N}$) with
a two-point one: 
\begin{equation}
(n_{\min},d_{n_{\min}}).
\end{equation}
What we propose now is that
\begin{lyxlist}{00.00.0000}
\item [{(A)}] there is no reason to stop at the smallest $n$ with $d_{n}>0$,
one can compute an entire function $n\mapsto d_{n}$ ($n=1,2,\ldots,N$),
the system's contextuality profile;
\item [{(B)}] one can do this for any well-constructed measure of contextuality;
\item [{(C)}] computing contextuality profiles for different measures can
be an informative way of comparing them.
\end{lyxlist}
We propose that a well-constructed contextuality measure should have
the following three properties:
\begin{enumerate}
\item for any noncontextual system its final-level degree of contextuality
is zero;
\item for any contextual system its final-level degree of contextuality
is positive;
\item its contextuality profile is a nondecreasing function of level.
\end{enumerate}
These requirements are obviously satisfied for the three measures
we are going to explore in the next section. However, it is worth
mentioning that some seemingly reasonable measures of contextuality
may fail them. Thus, in Ref. \citep{KujDzhMeasures} we describe a
measure abbreviated $\mathsf{CNT}_{1}$, which, as it turns out, may
produce decreasing contextuality profiles. One therefore should consider
this measure not well-constructed, and this is the reason we do not
include it in the analysis below.

\section{\label{sec:Three-measures-of}Three measures of contextuality}

The three measures of contextuality we are interested in are described
in detail in Ref. \citep{KujDzhMeasures}. Here we present their brief
characterization. 

The first measure is the already-mentioned ``hierarchical'' measure.
In this paper we call it the \emph{distance measure}. It is based
on the notion of distance between a system and a \emph{noncontextuality
polytope}. Consider all possible matrices of a given \emph{format}.
The latter is defined by the set of all questions $q$, the set of
all contexts $c$, and the relation $q\Yleft c$. Two systems of the
same format differ in the joint distributions within their corresponding
contexts.

\begin{figure}
\begin{centering}
\includegraphics[scale=0.3]{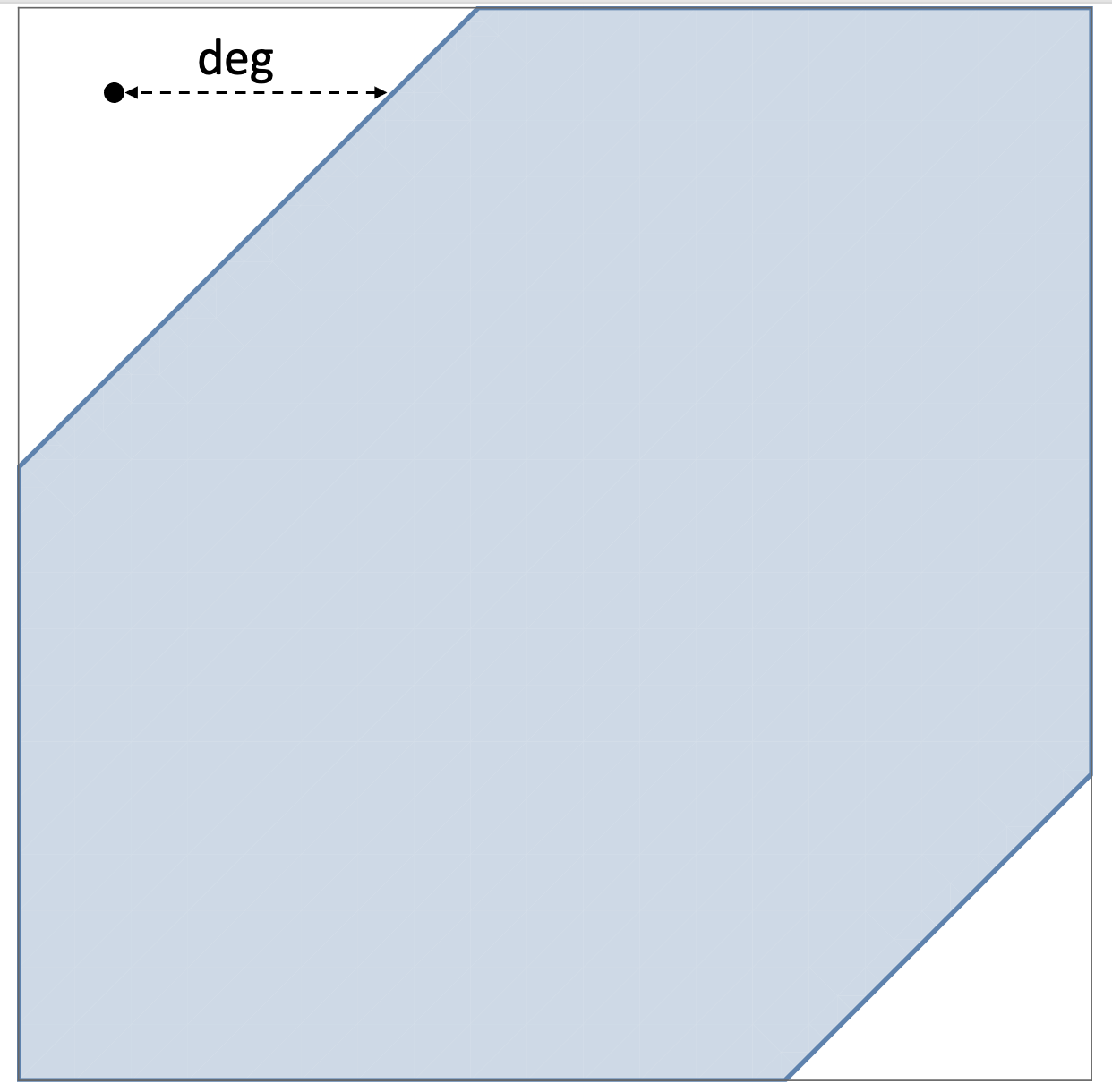}
\par\end{centering}
\caption{\label{fig:A-two-dimensional-projection}A two-dimensional projection
of a vector $\mathbf{v}^{*}$ and a noncontextuality polytope $\mathbb{V}$,
with the $L^{1}$-distance between them.}

\end{figure}

Let $\mathbb{V}$ be the set of all vectors $\mathbf{v}$ such that
equation (\ref{eq:noncontextuality equation}) has a solution:
\begin{equation}
\mathbb{V}=\left\{ \mathbf{v}:\mathbf{M}\,\mathbf{x}=\mathbf{v}\textnormal{ for some }\mathbf{x}\geq\mathbf{0}\right\} .
\end{equation}
It is known that in the space of $\mathbf{v}$-vectors $\mathbb{V}$
forms a polytope, which we call the noncontextuality polytope for
systems of a given format (with fixed individual distributions of
the variables). If the system we study (let's denote its vectorial
representation by $\mathbf{v}^{*}$) is contextual, then it falls
outside $\mathbb{V}$, and its distance from $\mathbb{V}$ can be
viewed as a measure of contextuality. When dealing with probability
distributions, the distance measure of choice is $L_{1}$, defined
by
\begin{equation}
L_{1}\left(\mathbf{a},\mathbf{b}\right)=\sum\left|a_{i}-b_{i}\right|,
\end{equation}
where the summation is over all dimensions of the vector space. Our
distance measure is 
\begin{equation}
\mathsf{deg}=L_{1}\left(\mathbf{\mathbf{v}^{*}},\mathbb{V}\right),
\end{equation}
the $L_{1}$-distance between the vector $\mathbf{v}^{*}$ and the
polytope $\mathbb{V}$, as shown in Figure \ref{fig:A-two-dimensional-projection}.
We denote this measure $\mathsf{CNT}_{2}$, following the nomenclature
adopted in several previous publications, e.g. \citep{KujDzhMeasures}. 

The next measure of contextuality is based on \emph{quasi-probabilities}
(positive and negative numbers that sum to 1). It is obtained by dropping
in the definition of noncontextuality (\ref{eq:noncontextuality equation})
the nonnegativity constraint $\mathbf{x}\geq\mathbf{0}$:
\begin{equation}
\mathbf{M}\:\mathbf{x}=\mathbf{v}\quad\left(\mathbf{1}^{\intercal}\mathbf{x}=1\right).\label{eq:CNT3 equality}
\end{equation}
This matrix equation is always solvable for $\mathbf{x}$, but some
of the components of a solution may be negative. Let $\left|\mathbf{x}\right|$
denote the vector of absolute values of the components of $\mathbf{x}$.
It is easy to see that
\begin{equation}
\mathbf{1}^{\intercal}\left|\mathbf{x}\right|\geq1,
\end{equation}
and the equality is achieved if and only if $\mathbf{x}$ contains
no negative components. Among all solutions $\mathbf{x}$ one can
always find some for which $\mathbf{1}^{\intercal}\left|\mathbf{x}\right|$
has the smallest possible value, and then
\begin{equation}
\mathsf{deg}=\mathbf{1}^{\intercal}\left|\mathbf{x}\right|-1\label{eq:CNT3 measure}
\end{equation}
is a measure of contextuality. We refer to it as the \emph{quasi-probability
measure} of contextuality, and denote it $\mathsf{CNT_{3}}$, as in
our previous publications.

For the third measure of contextuality, replace the equality in the
definition of noncontextuality (\ref{eq:noncontextuality equation})
with 
\begin{equation}
\mathbf{M}\,\mathbf{x}\leq\mathbf{v},\quad\mathbf{x}\geq\mathbf{0}\quad\left(\mathbf{1}^{\intercal}\mathbf{x}\leq1\right),\label{eq:CNTF inequality}
\end{equation}
where the inequalities are taken to hold component-wise. This inequality
always has solutions, and among them there are some with the maximal
value of $\mathbf{1}^{\intercal}\mathbf{x}$. Then 
\begin{equation}
\mathsf{deg}=1-\left(\mathbf{1}^{\intercal}\mathbf{x}\right)_{\max},\label{eq:CNTF measure}
\end{equation}
is a measure of contextuality, and it is called \emph{contextual fraction,
$\mathsf{CNTF}$.}

The contextual fraction and the quasi-probability measures were first
proposed by Abramsky and Brandenburger \citep{AB2011}. We later extended
them to also apply to disturbed systems (see Ref. \citep{book2026}
for details). Note that the vector $\mathbf{v}$ representing a system,
and therefore also the matrix $\mathbf{M}$, are generally different
for different measures of contextuality. For our three measures, they
are the same for $\mathsf{CNT}_{2}$ and $\mathsf{CNT}_{3}$ but different
for $\mathsf{CNTF}$ (for details, see Refs. \citep{book2026,KujDzhMeasures}).
Note also that when applied to undisturbed systems, $\mathsf{CNT}_{2}$
and $\mathsf{CNTF}$ produce the same contextuality values for complete
couplings and reduced couplings; however, $\mathsf{CNT}_{3}$ values
for complete and reduced couplings generally differ (which may be
viewed as a weakness of this measure).

Each of the three measures, $\mathsf{CNT}_{2}$, $\mathsf{CNT}_{3}$,
and $\mathsf{CNTF}$, can be, at least in principle, applied to systems
of any format. In particular, given a system $\mathcal{R}$, each
of them can be applied to all its level representations, 
\[
\mathcal{R}\left[1\right],\mathcal{R}\left[2\right],\ldots,\mathcal{R}\left[N\right],
\]
to form their respective profiles.

\section{\label{sec:Concatenated-systems}Concatenated systems}

Contextuality profiles can be studied in many ways, because, as all
functions, they can be characterized in many ways. Moreover, as should
be expected, their properties depend on the format of the systems
we choose. This paper being introductory, we focus here on one aspect
of contextuality profiles only: on comparing our three measures of
contextuality, $\mathsf{CNT_{2}}$, $\mathsf{CNT_{3}}$, and $\mathsf{CNTF}$,
on how fast the degree of contextuality tends to increase with its
level. Let us explain what we mean by this.

Suppose a measure $\mathsf{deg}$ produces a profile that changes
its value from $d_{n}$ to $d_{n+1}$ as one moves from level $n$
to level $n+1$; and suppose that the corresponding values for another
measure, $\mathsf{deg}'$, are $d'_{n}$ and $d'_{n+1}$. Both measures
are well-constructed, so
\begin{equation}
d_{n+1}\geq d_{n},d'_{n+1}\geq d'_{n}.
\end{equation}
However, it would not be informative to directly compare the numerical
values of $d_{n+1}-d_{n}$ and $d'_{n+1}-d'_{n}$ (unless one of these
differences is zero). The two measures are on completely different
scales, so we may be comparing meters to grams, or even worse, meters
to decibels. The same reasoning applies, of course, to their ratios,
differences of their cubes, or other measures of incrementation. We
need to find a way to consider the increase from $d_{n}$ to $d_{n+1}$
and from $d'_{n}$ to $d'_{n+1}$ intrinsically, within their respective
scales.

How can this be done? The increase from $d_{n}$ to $d_{n+1}$ occurs
because the system $\mathcal{R}\left[n+1\right]$ contains $\left(n+1\right)$-tuples
of variables, in addition to the $k$-tuples of variables with $k\leq n$
contained in $\mathcal{R}\left[n\right]$. The degree of contextuality
brought in by these $\left(n+1\right)$-tuples somehow combines with
the contextuality present in $\mathcal{R}\left[n\right]$ to produce
$d_{n+1}$. If we had a way of measuring the contextuality $\Delta_{n+1}$
brought in by these $\left(n+1\right)$-tuples only, then we would
be able to compare $d_{n}+\Delta_{n+1}$ to $d_{n+1}$:
\begin{lyxlist}{00.00.0000}
\item [{(1)}] $d_{n}+\Delta_{n+1}<d_{n+1}$ (superadditive increment) ,
\item [{(2)}] $d_{n}+\Delta_{n+1}=d_{n+1}$ (additive increment),
\item [{(3)}] $d_{n}+\Delta_{n+1}>d_{n+1}$ (subadditive increment),
\item [{(4)}] $d_{n}=d_{n+1}$ (plateau).
\end{lyxlist}
But is there a way to find $\Delta_{n+1}$ independently of $d_{n+1}$?
We propose one such way as follows.

Consider two systems,
\begin{equation}
\begin{array}{|c|c|c|c||c|}
\hline A_{1}^{1} & A_{2}^{1} & \cdots & A_{n}^{1} & c=1\\
\hline A_{1}^{2} & A_{2}^{2} & \cdots & A_{n}^{2} & 2\\
\hline \vdots & \vdots & \ddots & \vdots & \vdots\\
\hline A_{1}^{s} & A_{2}^{s} & \cdots & A_{n}^{s} & s\\
\hline\hline q=1 & 2 & \cdots & n & \mathcal{A}
\\\hline \end{array}
\end{equation}
and
\[
\begin{array}{|c|c|c|c|c||c|}
\hline B_{1}^{1} & B_{2}^{1} & \cdots & B_{n}^{1} & B_{n+1}^{1} & c=1'\\
\hline B_{1}^{2} & B_{2}^{2} & \cdots & B_{n}^{2} & B_{n+1}^{2} & 2'\\
\hline \vdots & \vdots & \ddots & \vdots & \vdots & \vdots\\
\hline B_{1}^{t} & B_{2}^{t} & \cdots & B_{n}^{t} & B_{n+1}^{t} & t'\\
\hline\hline q=1' & 2' & \cdots & n' & \left(n+1\right)' & \mathcal{B}
\\\hline \end{array}\:,
\]
where some of the variables shown can be constants or empty cells.
Let system $\mathcal{A}$ have a contextuality profile 
\begin{equation}
\begin{array}{c|c|c|c|c}
k & 1 & 2 & \cdots & n\\
\hline \mathsf{deg}\mathcal{A}\left[k\right] & 0 & d_{2} & \cdots & d_{n}\geq0
\end{array}.
\end{equation}
For system $\mathcal{B}$, let us assume that its contextuality profile
is 
\begin{equation}
\begin{array}{c|c|c|c|c|c}
k & 1 & 2 & \cdots & n & n+1\\
\hline \mathsf{deg}\mathcal{B}\left[k\right] & 0 & 0 & \cdots & 0 & \Delta_{n+1}\geq0
\end{array}.
\end{equation}
That is, this system is noncontextual at all levels except for the
last one. Let us concatenate these two systems into a larger system
as shown:
\begin{equation}
\begin{array}{|cccc||ccccc||c|}
\hline A_{1}^{1} & A_{2}^{1} & \cdots & A_{n}^{1} &  &  &  &  &  & c=1\\
A_{1}^{2} & A_{2}^{2} & \cdots & A_{n}^{2} &  &  &  &  &  & 2\\
\vdots & \vdots & \ddots & \vdots &  &  &  &  &  & \vdots\\
A_{1}^{s} & A_{2}^{s} & \cdots & A_{n}^{s} &  &  &  &  &  & s\\
\hline\hline  &  &  &  & B_{1}^{1} & B_{2}^{1} & \cdots & B_{n}^{1} & B_{n+1}^{1} & 1'\\
 &  &  &  & B_{1}^{2} & B_{2}^{2} & \cdots & B_{n}^{2} & B_{n+1}^{2} & 2'\\
 &  &  &  & \vdots & \vdots & \ddots & \vdots & \vdots & \vdots\\
 &  &  &  & B_{1}^{t} & B_{2}^{t} & \cdots & B_{n}^{t} & B_{n+1}^{t} & t'\\
\hline\hline q=1 & 2 & \cdots & n & 1' & 2' & \cdots & n' & \left(n+1\right)' & \mathcal{A}*\mathcal{B}
\\\hline \end{array}\:.
\end{equation}
Note that the sets of both questions and contexts of the two subsystems
are completely disjoint. This means that a coupling of $\mathcal{A}*\mathcal{B}$
can be constructed as separate couplings for $\mathcal{A}$ and for
$\mathcal{B}$, with their joint distribution defined arbitrarily
(in particular, they can always be treated as independent events).
In other words, for all $k\leq n$, since $\mathcal{B}\left[k\right]$
is noncontextual, the contextuality of $\left(\mathcal{A}*\mathcal{B}\right)\left[k\right]$
is determined by $\mathcal{A}\left[k\right]$ alone. It is natural
to expect then that, for all $k\leq n$,
\begin{equation}
\mathsf{deg}\left(\mathcal{A}*\mathcal{B}\right)\left[k\right]=\mathsf{deg}\left(\mathcal{A}\right)\left[k\right].
\end{equation}
This can even be added as a fourth requirement for a well-constructed
measure of contextuality, in addition to the three requirements listed
at the end of Section \ref{sec:Profiles-of-contextuality}. Whether
we do this or not, this property holds for all three measures $\mathsf{CNT_{2}}$,
$\mathsf{CNT_{3}}$, and $\mathsf{CNTF}$. Consequently, for all of
them we have
\begin{equation}
\begin{array}{c|c|c|c|c}
k & 1 & 2 & \cdots & n\\
\hline \mathsf{deg}\mathcal{A}\left[k\right] & 0 & d_{2} & \cdots & d_{n}\\
\hline \mathsf{deg}\mathcal{B}\left[k\right] & 0 & 0 & \cdots & 0\\
\hline \mathsf{deg}\left(\mathcal{A}*\mathcal{B}\right)\left[k\right] & 0 & d_{2} & \cdots & d_{n}
\end{array}.
\end{equation}
At the next, $\left(n+1\right)$st level we get

\begin{equation}
\begin{array}{c|c|c|c|c|c}
k & 1 & 2 & \cdots & n & n+1\\
\hline \mathsf{deg}\mathcal{A}\left[k\right] & 0 & d_{2} & \cdots & d_{n} & d_{n}\\
\hline \mathsf{deg}\mathcal{B}\left[k\right] & 0 & 0 & \cdots & 0 & \Delta_{n+1}\\
\hline \mathsf{deg}\left(\mathcal{A}*\mathcal{B}\right)\left[k\right] & 0 & d_{2} & \cdots & d_{n} & d_{n+1}
\end{array}.
\end{equation}
The reason $d_{n}$ repeats at level $n+1$ for $\mathsf{deg}\mathcal{A}\left[k\right]$
is that $\mathcal{A}\left[n+1\right]=\mathcal{A}\left[n\right]$.

Clearly, we now have what we have aimed at: the possibility to compare
$d_{n+1}$ and $d_{n}+\Delta_{n+1}$, in order to determine if the
combination of $d_{n}$ and $\Delta_{n+1}$ by the measure $\mathsf{deg}$
is additive, superadditive, or subadditive (including the plateau
case, d$_{n+1}=d_{n}$). 

\section{\label{sec:Contextuality-profiles-for}Contextuality profiles for
the three measures}

\begin{figure}
\begin{centering}
\includegraphics[scale=0.35]{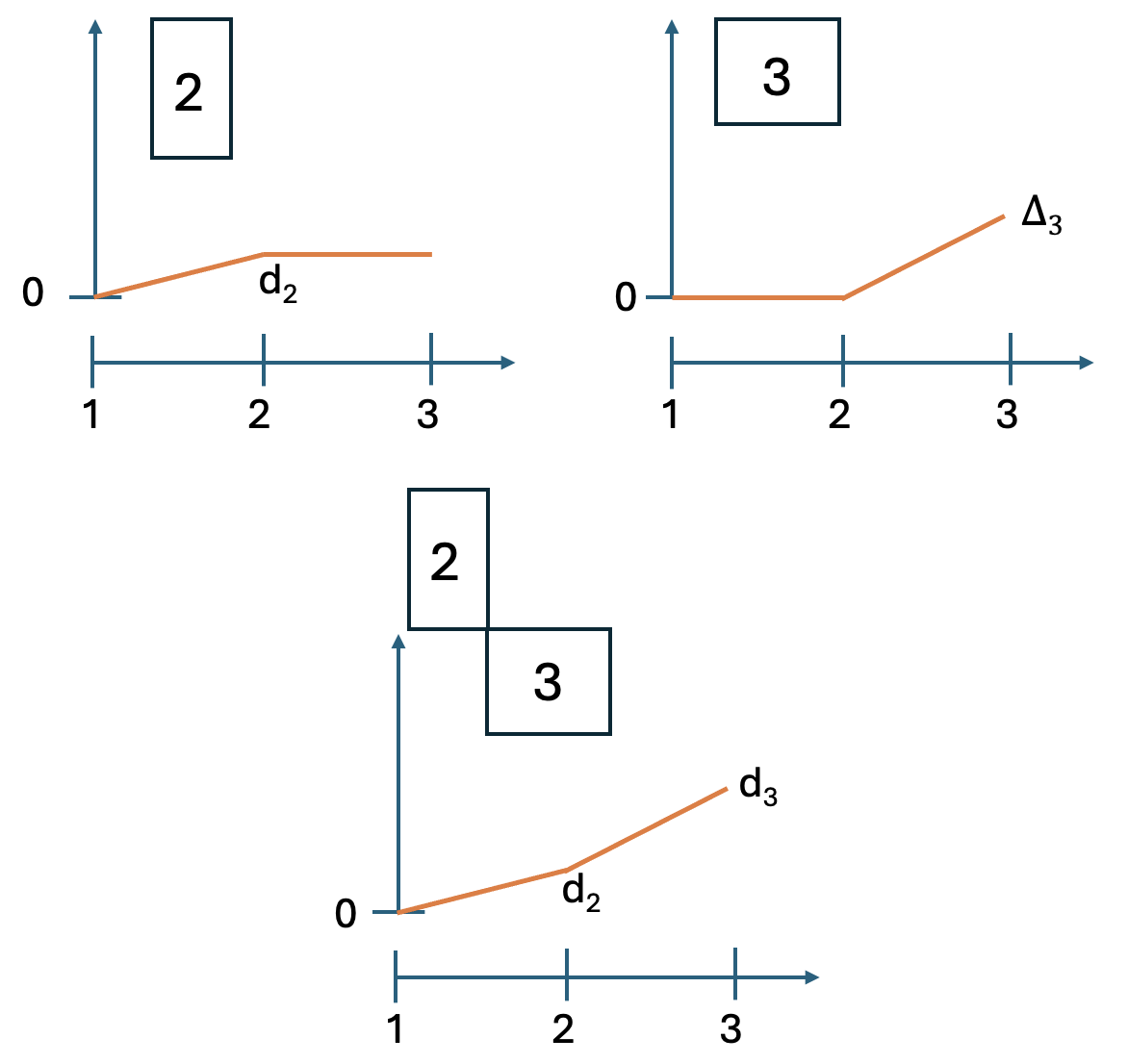}
\par\end{centering}
\caption{\label{fig:Contextuality-profiles-for}Contextuality profiles for
the method of concatenated systems, $n=2$. The boxes represent the
systems being concatenated, with the numbers in them indicating their
final level of contextuality. Symbols attached to the curves indicate
contextuality values.}
\end{figure}

\begin{figure}
\begin{centering}
\includegraphics[scale=0.35]{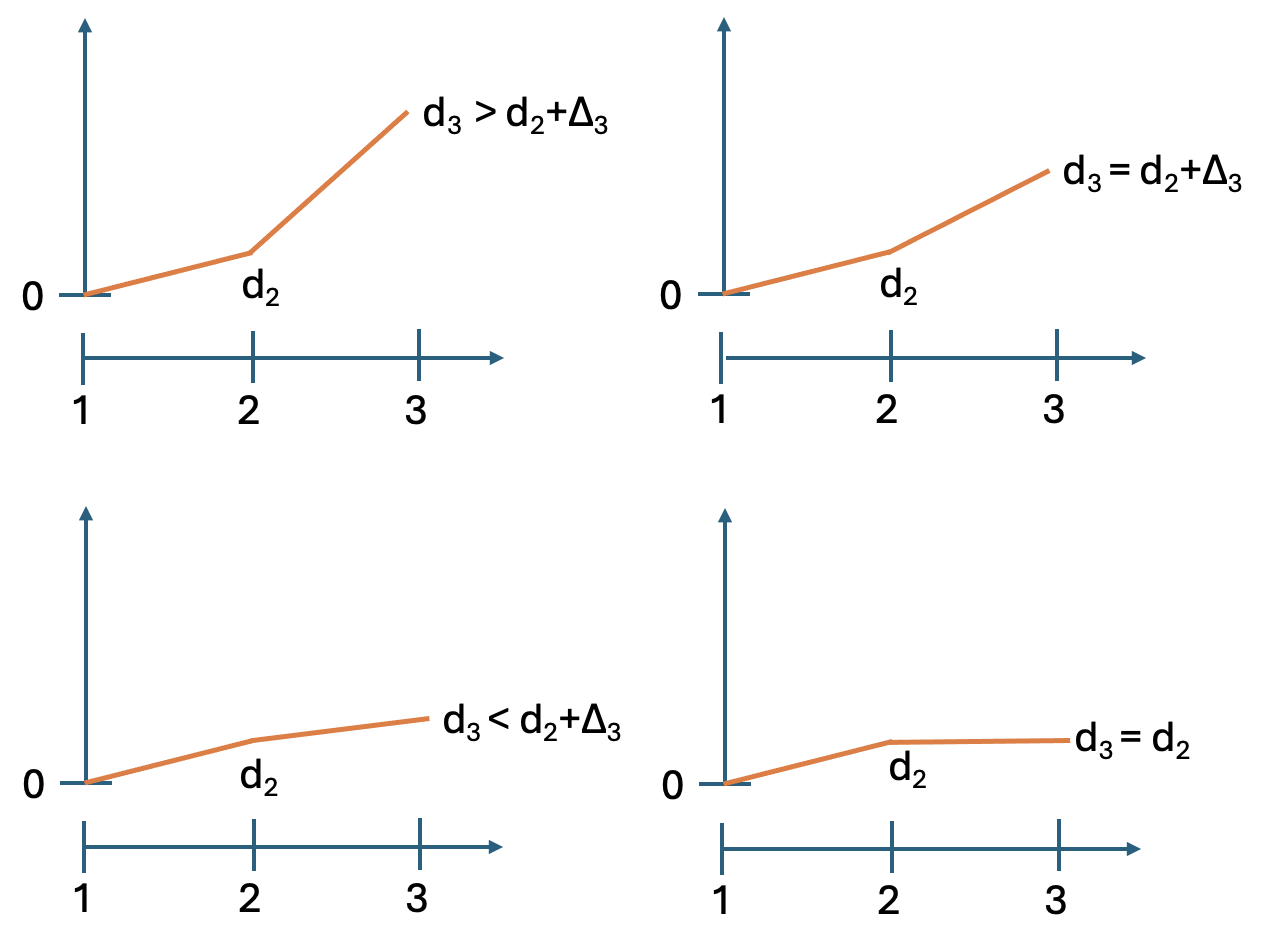}
\par\end{centering}
\caption{\label{fig:Four-possible-types}Four possible types of the contextuality
profiles for concatenated systems ($n=2$): superadditive (top left
panel), additive (top right), subadditive (bottom left), and, as the
extreme case of subadditivity, plateau (bottom right).}
\end{figure}

We implement the method presented in the previous section using its
simplest special case: with $n=2$. Figures \ref{fig:Contextuality-profiles-for}
and \ref{fig:Four-possible-types} illustrate the logic and the possible
types of contextuality profiles for this special case. 

We chose the formats for systems $\mathcal{A}$ and $\mathcal{B}$
as shown,

\begin{equation}
\begin{array}{|c|c|c||c|}
\hline A_{1}^{1} & A_{2}^{1} &  & 1\\
\hline  & A_{2}^{2} & A_{3}^{2} & 2\\
\hline A_{1}^{3} &  & A_{3}^{3} & 3\\
\hline\hline 1 & 2 & 3 & \mathcal{A}
\\\hline \end{array}\:,\qquad\begin{array}{|c|c|c|c||c|}
\hline B_{1}^{1} & B_{2}^{1} & B_{3}^{1} &  & 1'\\
\hline  & B_{2}^{2} & B_{3}^{2} & B_{4}^{2} & 2'\\
\hline B_{1}^{3} &  & B_{3}^{3} & B_{4}^{3} & 3'\\
\hline\hline 1' & 2' & 3' & 4' & \mathcal{B}
\\\hline \end{array}\:,
\end{equation}
with all variables being dichotomous (say, $\pm1$). The format of
the concatenated system then acquires the form
\begin{equation}
\begin{array}{|ccc|cccc||c|}
\hline A_{1}^{1} & A_{2}^{1} &  &  &  &  &  & 1\\
 & A_{2}^{2} & A_{3}^{2} &  &  &  &  & 2\\
A_{1}^{3} &  & A_{3}^{3} &  &  &  &  & 3\\
\hline  &  &  & B_{1}^{1} & B_{2}^{1} & B_{3}^{1} &  & 1'\\
 &  &  &  & B_{2}^{2} & B_{3}^{2} & B_{4}^{2} & 2'\\
 &  &  & B_{1}^{3} &  & B_{3}^{3} & B_{4}^{3} & 3'\\
\hline\hline 1 & 2 & 3 & 1' & 2' & 3' & 4' & \mathcal{A*\mathcal{B}}
\\\hline \end{array}\:.
\end{equation}
The systems we explored were obtained by specifying the joint distribution
of the variables in the systems $\mathcal{A}$ and $\mathcal{B}$.
Each contextuality profile shown below has symbols attached to it,
referring to the systems whose detailed specifications are given in
Appendix A. 

\begin{figure}
\begin{centering}
\includegraphics[scale=0.35]{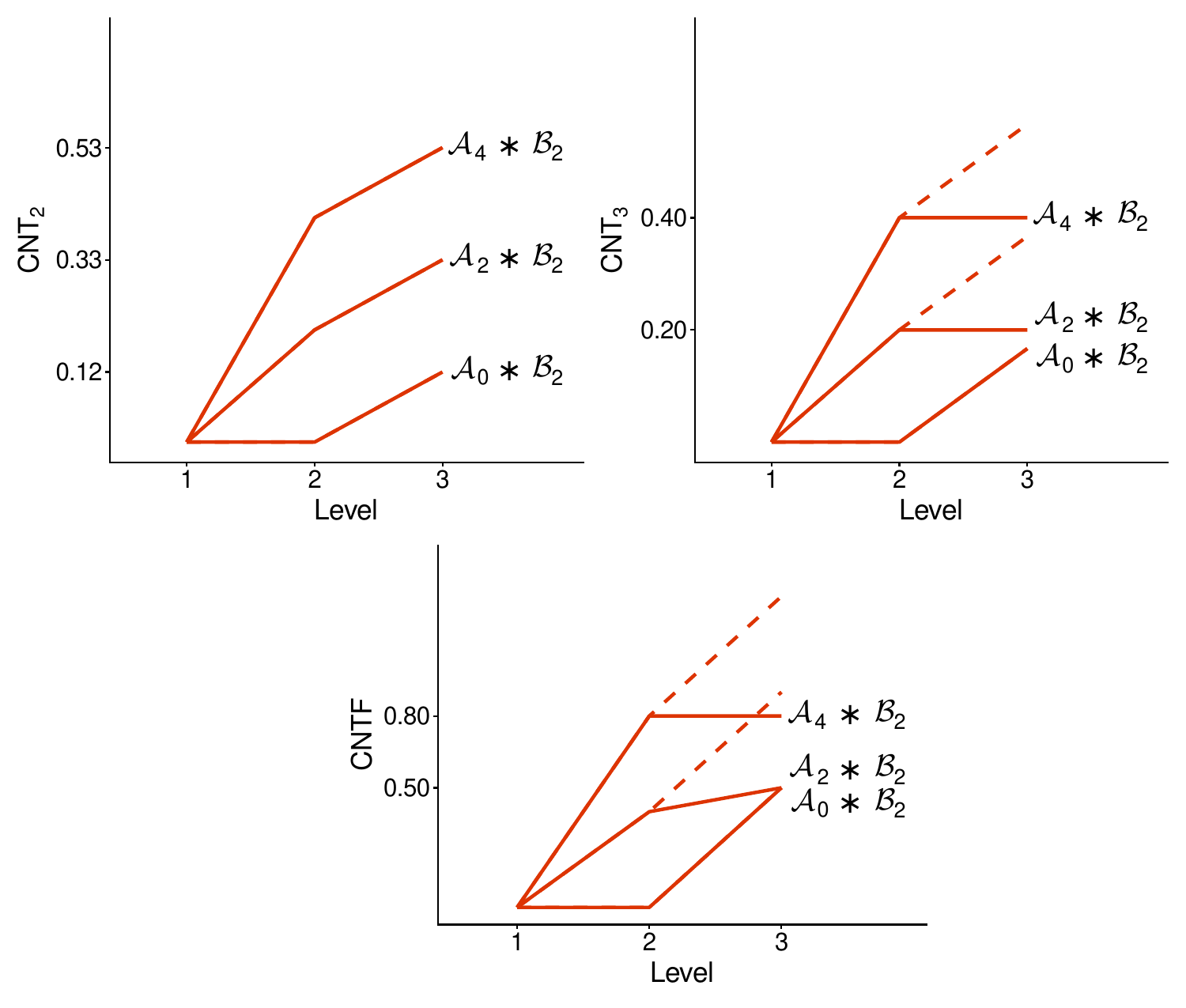}
\par\end{centering}
\caption{\label{fig:Contextuality-profiles-for-even-1}Contextuality profiles
for a selection of undisturbed concatenated systems. Symbols $\mathcal{A}$
and $\mathcal{B}$ with indices refer to $\mathcal{A}$- and $\mathcal{B}$-subsystems,
respectively (as specified in Appendix A). The dashed lines attached
to each profile show the increment from $d_{2}$ to $d_{2}+\Delta_{3}$:
if it is above the corresponding segment of the profile we have subadditivity,
and when the dashed line is not seen (coincides with the segment)
we have additivity.}

\end{figure}

\begin{figure}
\begin{centering}
\includegraphics[scale=0.35]{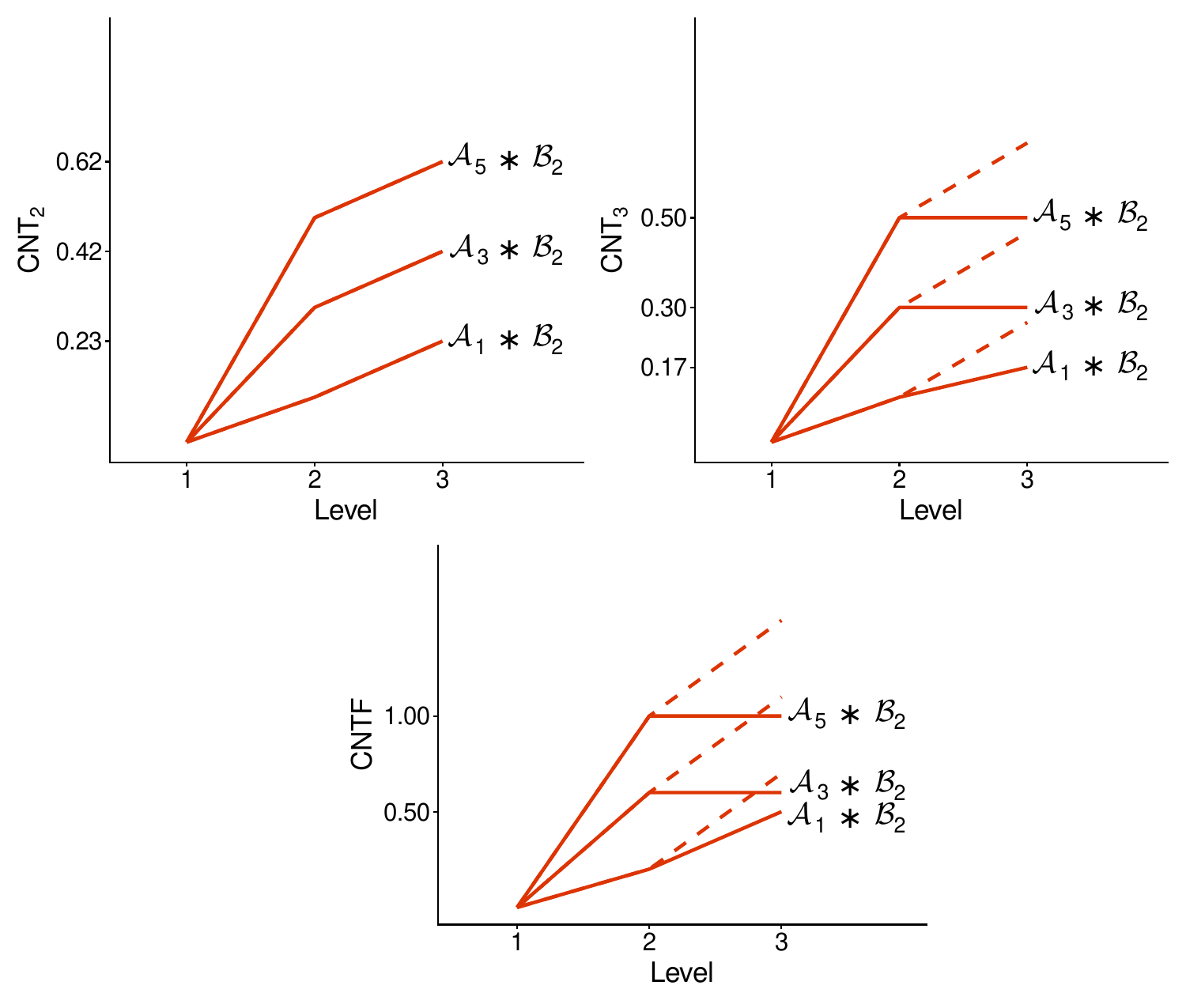}
\par\end{centering}
\caption{\label{fig:Contextuality-profiles-for-odd-2}The same as in Figure
\ref{fig:Contextuality-profiles-for-even-1}, for another selection
of the $\mathcal{A}$-subsystems.}
\end{figure}

Figures \ref{fig:Contextuality-profiles-for-even-1}-\ref{fig:Contextuality-profiles-for-odd-2}
show the contextuality profiles for a selection of undisturbed concatenated
systems. We see that the measure $\mathsf{CNT_{2}}$ shows precise
additivity, while both $\mathsf{CNT_{3}}$ and $\mathsf{CNTF}$ are
subadditive. The subadditivity in these measures, especially in $\mathsf{CNT_{3}}$,
is often extreme, resulting in a plateau in most cases shown.

\begin{figure}
\begin{centering}
\includegraphics[scale=0.35]{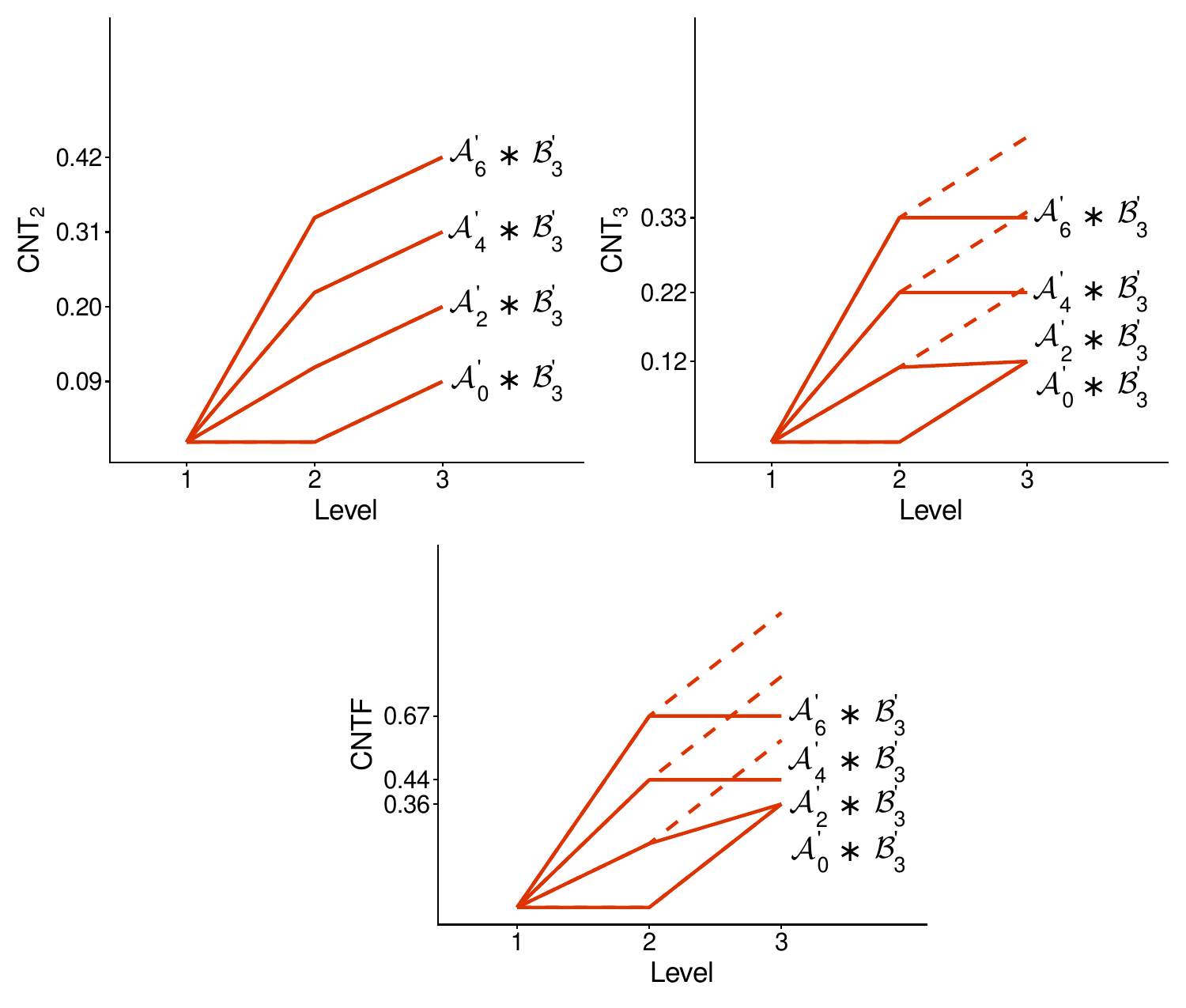}
\par\end{centering}
\caption{\label{fig:Contextuality-profiles-disturbed}Contextuality profiles
for a selection of disturbed $\mathcal{A}$-subsystems concatenated
with a disturbed subsystem $\mathcal{B}_{3}$. The rest is the same
as in Figure \ref{fig:Contextuality-profiles-for-even-1}.}
\end{figure}

\begin{figure}
\begin{centering}
\includegraphics[scale=0.35]{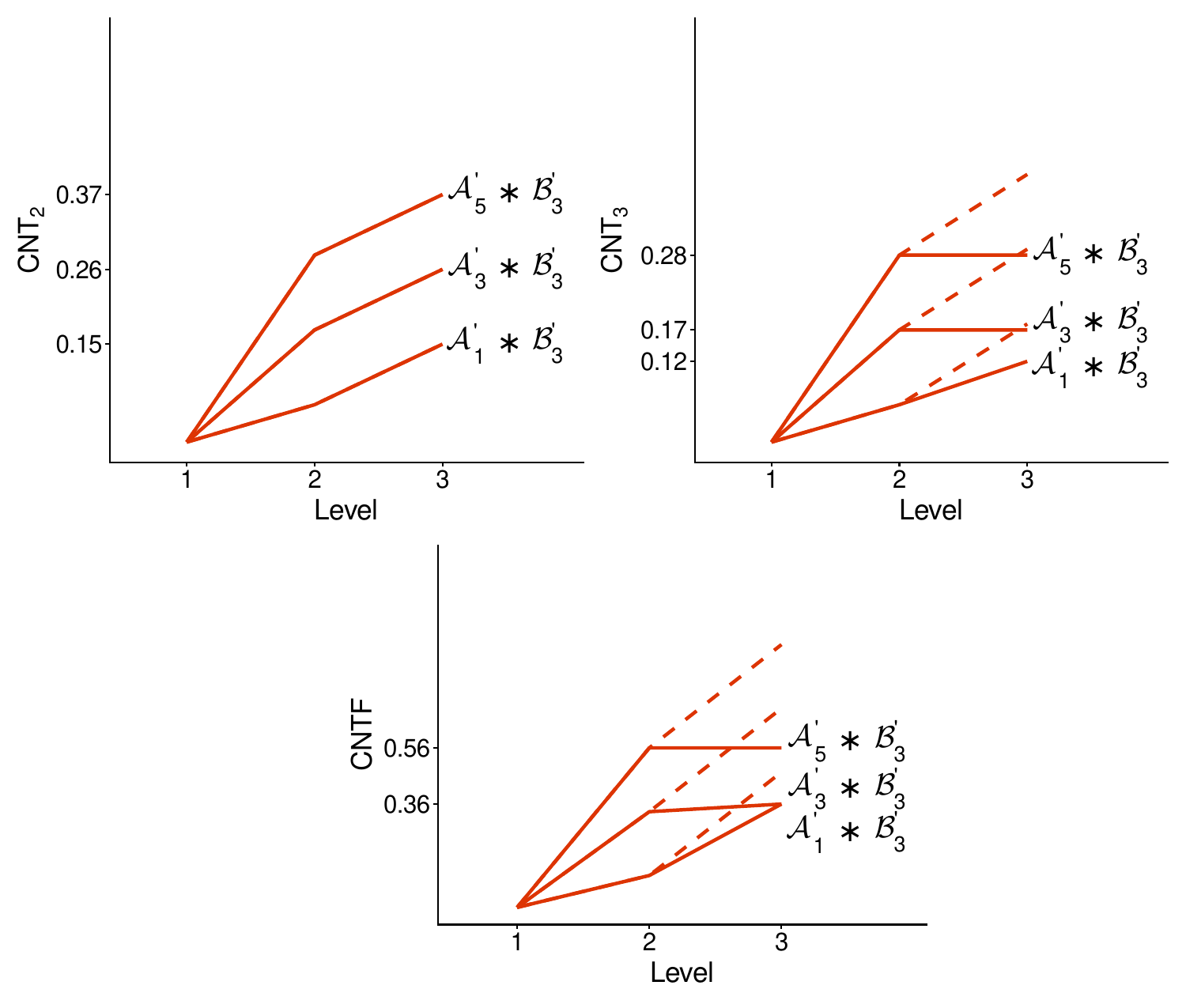}
\par\end{centering}
\caption{\label{fig:Contextuality-profiles-disturbed-3-2}The same as in Figure
\ref{fig:Contextuality-profiles-disturbed} but for another selection
of $\mathcal{A}$-subsystems concatenated with a disturbed subsystem
$\mathcal{B}_{3}$.}
\end{figure}

There is no qualitative difference between the profiles of the undisturbed
and disturbed concatenated systems. Figures \ref{fig:Contextuality-profiles-disturbed}-\ref{fig:Contextuality-profiles-disturbed-3-2}
for a selection of disturbed systems exhibit the same pattern as in
Figures \ref{fig:Contextuality-profiles-for-even-1}-\ref{fig:Contextuality-profiles-for-odd-2}. 

\begin{figure}
\begin{centering}
\includegraphics[scale=0.35]{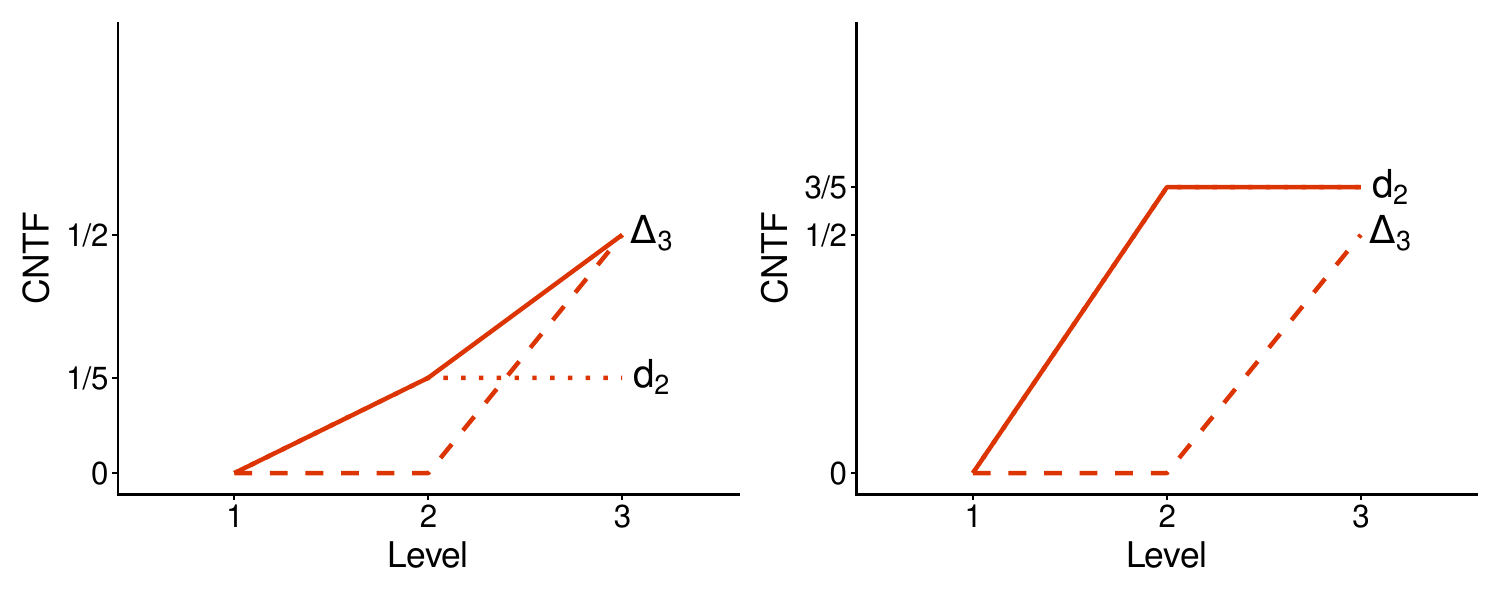}
\par\end{centering}
\caption{\label{fig:maxCNTF}$\mathsf{CNTF}$ profiles for a selection of concatenated
systems ($\mathcal{A}_{1}\ast\mathcal{B}_{2}$ left and $\mathcal{A}_{3}\ast\mathcal{B}_{2}$
right). The dashed lines represent the $\mathsf{CNTF}$ profiles for
the systems' $\mathcal{B}$-parts. The dotted lines represent the
$\mathsf{CNTF}$ profiles for the system's $\mathcal{A}$-parts (invisible
if it coincides with a system's profile).}
\end{figure}

\begin{figure}
\begin{centering}
\includegraphics[scale=0.35]{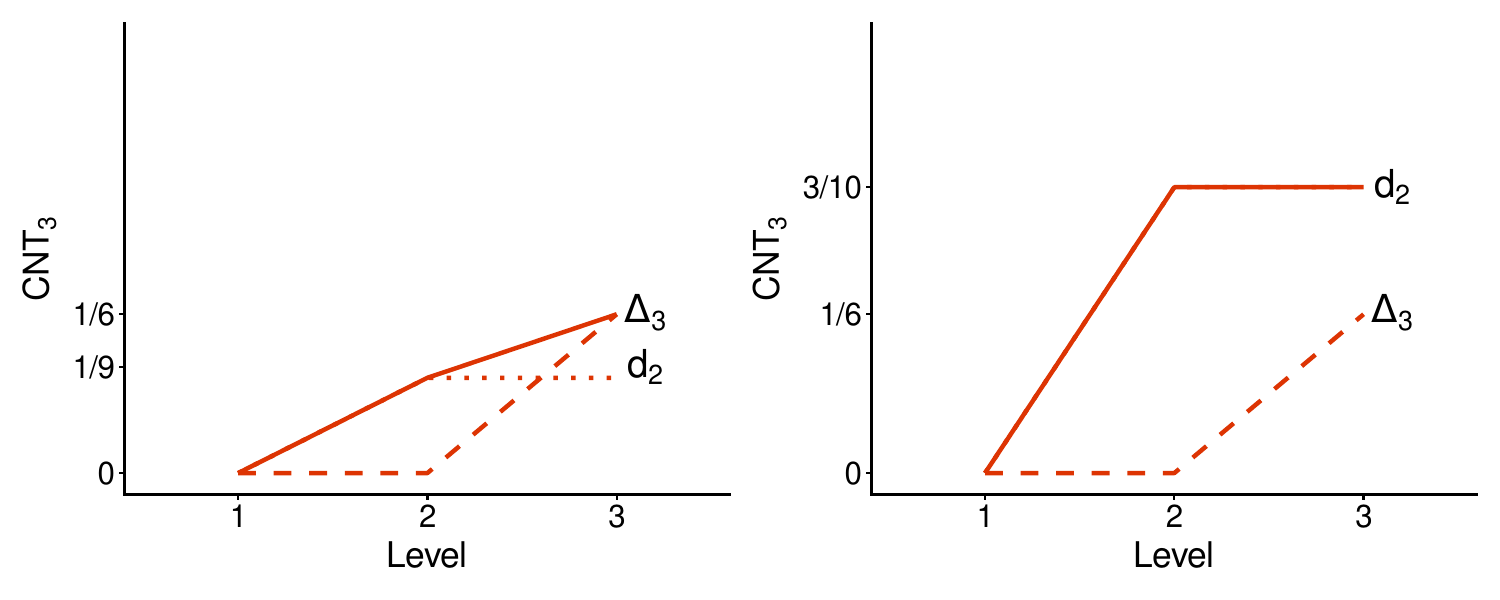}
\par\end{centering}
\caption{\label{fig:maxCNT3}The same as in Figure \ref{fig:maxCNTF} but for
$\mathsf{CNT_{3}}$.}
\end{figure}

Now that the subadditivity of the contextuality profiles for $\mathsf{CNT_{3}}$
and $\mathsf{CNTF}$ has been observed, can we determine its cause?
It turns out we can. Figures \ref{fig:maxCNTF} and \ref{fig:maxCNT3}
exhibit the contextuality profiles for $\mathsf{CNT_{3}}$ and $\mathsf{CNTF}$
with the superimposed profiles of the individual $\mathcal{A}$- and
$\mathcal{B}$-subsystems. One can see that $d_{3}$ coincides with
$\Delta_{3}$ if the latter exceeds $d_{2}$; otherwise $d_{3}$ remains
on the level of $d_{2}$. In other words, for both $\mathsf{CNT_{3}}$
and $\mathsf{CNTF}$ profiles shown, we have the rule of maximum:
\begin{equation}
d_{3}=\max\left(d_{2},\Delta_{3}\right).\label{eq:maximum rule}
\end{equation}
The subadditivity therefore is the consequence of
\[
\max\left(d_{2},\Delta_{3}\right)\leq d_{2}+\Delta_{3}.
\]

Tables \ref{tab:A-table-showing} and \ref{tab:A-table-showing-1}
provides an illustration of the addition rule for $\mathsf{CNT_{2}}$
and the rule of maximum for $\mathsf{CNT_{3}}$ and $\mathsf{CNTF}$
using larger selections of subsystems $\mathcal{A}$ and $\mathcal{B}$
than in our figures. 

\begin{table}[h]
\begin{centering}
\begin{tabular}{|cc|cccccc|}
\cline{3-8}
\multicolumn{1}{c}{} & \multirow{2}{*}{$\mathsf{CNT_{2}}$} & $\mathcal{A}_{0}$ & $\mathcal{A}_{1}$ & $\mathcal{A}_{2}$ & $\mathcal{A}_{3}$ & $\mathcal{A}_{4}$ & $\mathcal{A}_{5}$\tabularnewline
\multicolumn{1}{c}{} &  & 0 & $\nicefrac{1}{10}$ & $\nicefrac{1}{5}$ & $\nicefrac{3}{10}$ & $\nicefrac{2}{5}$ & $\nicefrac{1}{2}$\tabularnewline
\hline 
$\mathcal{B}_{1}$ & $\nicefrac{1}{24}$ & $\nicefrac{1}{24}$ & $\nicefrac{17}{120}$ & $\nicefrac{29}{120}$ & $\nicefrac{41}{120}$ & $\nicefrac{53}{120}$ & $\nicefrac{13}{24}$\tabularnewline
$\mathcal{B}_{2}$ & $\nicefrac{1}{8}$ & $\nicefrac{1}{8}$ & $\nicefrac{9}{40}$ & $\nicefrac{13}{40}$ & $\nicefrac{17}{40}$ & $\nicefrac{21}{40}$ & $\nicefrac{5}{8}$\tabularnewline
\hline 
\multicolumn{1}{c}{} & \multirow{2}{*}{$\mathsf{CNT_{3}}$} & $\mathcal{A}_{0}$ & $\mathcal{A}_{1}$ & $\mathcal{A}_{2}$ & $\mathcal{A}_{3}$ & $\mathcal{A}_{4}$ & $\mathcal{A}_{5}$\tabularnewline
\multicolumn{1}{c}{} &  & 0 & $\nicefrac{1}{10}$ & $\nicefrac{1}{5}$ & $\nicefrac{3}{10}$ & $\nicefrac{2}{5}$ & $\nicefrac{1}{2}$\tabularnewline
\hline 
$\mathcal{B}_{1}$ & $\nicefrac{1}{18}$ & $\nicefrac{1}{18}$ & $\nicefrac{1}{10}$ & $\nicefrac{1}{5}$ & $\nicefrac{3}{10}$ & $\nicefrac{2}{5}$ & $\nicefrac{1}{2}$\tabularnewline
$\mathcal{B}_{2}$ & $\nicefrac{1}{6}$ & $\nicefrac{1}{6}$ & $\nicefrac{1}{6}$ & $\nicefrac{1}{5}$ & $\nicefrac{3}{10}$ & $\nicefrac{2}{5}$ & $\nicefrac{1}{2}$\tabularnewline
\hline 
\multicolumn{1}{c}{} & \multirow{2}{*}{$\mathsf{CNTF}$} & $\mathcal{A}_{0}$ & $\mathcal{A}_{1}$ & $\mathcal{A}_{2}$ & $\mathcal{A}_{3}$ & $\mathcal{A}_{4}$ & $\mathcal{A}_{5}$\tabularnewline
\multicolumn{1}{c}{} &  & 0 & $\nicefrac{1}{5}$ & $\nicefrac{2}{5}$ & $\nicefrac{3}{5}$ & $\nicefrac{4}{5}$ & $1$\tabularnewline
\hline 
$\mathcal{B}_{1}$ & $\nicefrac{1}{6}$ & $\nicefrac{1}{6}$ & $\nicefrac{1}{5}$ & $\nicefrac{2}{5}$ & $\nicefrac{3}{5}$ & $\nicefrac{4}{5}$ & $1$\tabularnewline
\cline{3-8}
$\mathcal{B}_{2}$ & $\nicefrac{1}{2}$ & $\nicefrac{1}{2}$ & $\nicefrac{1}{2}$ & $\nicefrac{1}{2}$ & $\nicefrac{3}{5}$ & $\nicefrac{4}{5}$ & $1$\tabularnewline
\hline 
\end{tabular}
\par\end{centering}
\caption{\label{tab:A-table-showing}Contextuality value $d_{3}$ of undisturbed
concatenated systems for the measures $\mathsf{CNT_{2}}$, $\mathsf{CNT_{3}}$,
and $\mathsf{CNTF}.$ The value of $d_{3}$ for $\mathcal{A}_{i}*\mathcal{B}_{j}$
is in the intersection of column $\mathcal{A}_{i}$ and row $\mathcal{B}_{j}$.
The corresponding values of $d_{2}$ for $\mathcal{A}_{i}$ and $\Delta_{3}$
for $\mathcal{B}_{j}$ are shown, respectively, in the row and the
column containing the measure's name. Observe that for $\mathsf{CNT_{2}}$,
$d_{3}$ is the sum of the corresponding values of $d_{2}$ and $\Delta_{3}$;
and for both $\mathsf{CNT_{3}}$ and $\mathsf{CNTF}$, $d_{3}$ is
the larger of the corresponding values of $d_{2}$ and $\Delta_{3}$.
(Note that for the $\mathcal{A}$-subsystems, $\mathsf{CNT_{2}=\mathsf{CNT_{3}}=\mathsf{\frac{1}{2}CNTF}}$,
as it was previously established for all cyclic systems \citep{CamilloCervantes 2024}). }
\end{table}

\begin{table}[h]
\begin{centering}
\begin{tabular}{|cc|ccccccc|}
\cline{3-9}
\multicolumn{1}{c}{} & \multirow{2}{*}{$\mathsf{CNT_{2}}$} & $\mathcal{A}'_{0}$ & $\mathcal{A}'_{1}$ & $\mathcal{A}'_{2}$ & $\mathcal{A}'_{3}$ & $\mathcal{A}'_{4}$ & $\mathcal{A}'_{5}$ & $\mathcal{A}'_{6}$\tabularnewline
\multicolumn{1}{c}{} &  & 0 & $\nicefrac{1}{18}$ & $\nicefrac{1}{9}$ & $\nicefrac{1}{6}$ & $\nicefrac{2}{9}$ & $\nicefrac{5}{18}$ & $\nicefrac{3}{9}$\tabularnewline
\hline 
$\mathcal{B}'_{1}$ & $\nicefrac{1}{100}$ & $\nicefrac{1}{100}$ & $\nicefrac{59}{900}$ & $\nicefrac{109}{900}$ & $\nicefrac{159}{900}$ & $\nicefrac{209}{900}$ & $\nicefrac{259}{900}$ & $\nicefrac{309}{900}$\tabularnewline
$\mathcal{B}'_{2}$ & $\nicefrac{5}{100}$ & $\nicefrac{5}{100}$ & $\nicefrac{19}{180}$ & $\nicefrac{29}{180}$ & $\nicefrac{13}{60}$ & $\nicefrac{49}{180}$ & $\nicefrac{59}{180}$ & $\nicefrac{23}{180}$\tabularnewline
$\mathcal{B}'_{3}$ & $\nicefrac{9}{100}$ & $\nicefrac{9}{100}$ & $\nicefrac{131}{900}$ & $\nicefrac{181}{900}$ & $\nicefrac{77}{300}$ & $\nicefrac{281}{900}$ & $\nicefrac{331}{900}$ & $\nicefrac{127}{300}$\tabularnewline
\hline 
\multicolumn{1}{c}{} & \multirow{2}{*}{$\mathsf{CNT_{3}}$} & $\mathcal{A}'_{0}$ & $\mathcal{A}'_{1}$ & $\mathcal{A}'_{2}$ & $\mathcal{A}'_{3}$ & $\mathcal{A}'_{4}$ & $\mathcal{A}'_{5}$ & $\mathcal{A}'_{6}$\tabularnewline
\multicolumn{1}{c}{} &  & 0 & $\nicefrac{1}{18}$ & $\nicefrac{1}{9}$ & $\nicefrac{1}{6}$ & $\nicefrac{2}{9}$ & $\nicefrac{5}{18}$ & $\nicefrac{3}{9}$\tabularnewline
\hline 
$\mathcal{B}'_{1}$ & $\nicefrac{2}{150}$ & $\nicefrac{2}{150}$ & $\nicefrac{1}{18}$ & $\nicefrac{1}{9}$ & $\nicefrac{1}{6}$ & $\nicefrac{2}{9}$ & $\nicefrac{5}{18}$ & $\nicefrac{3}{9}$\tabularnewline
$\mathcal{B}'_{2}$ & $\nicefrac{1}{15}$ & $\nicefrac{1}{15}$ & $\nicefrac{1}{15}$ & $\nicefrac{1}{9}$ & $\nicefrac{1}{6}$ & $\nicefrac{2}{9}$ & $\nicefrac{5}{18}$ & $\nicefrac{3}{9}$\tabularnewline
$\mathcal{B}'_{3}$ & $\nicefrac{3}{25}$ & $\nicefrac{3}{25}$ & $\nicefrac{3}{25}$ & $\nicefrac{3}{25}$ & $\nicefrac{1}{6}$ & $\nicefrac{2}{9}$ & $\nicefrac{5}{18}$ & $\nicefrac{3}{9}$\tabularnewline
\hline 
\multicolumn{1}{c}{} & \multirow{2}{*}{$\mathsf{CNTF}$} & $\mathcal{A}'_{0}$ & $\mathcal{A}'_{1}$ & $\mathcal{A}'_{2}$ & $\mathcal{A}'_{3}$ & $\mathcal{A}'_{4}$ & $\mathcal{A}'_{5}$ & $\mathcal{A}'_{6}$\tabularnewline
\multicolumn{1}{c}{} &  & 0 & $\nicefrac{1}{9}$ & $\nicefrac{2}{9}$ & $\nicefrac{1}{3}$ & $\nicefrac{4}{9}$ & $\nicefrac{5}{9}$ & $\nicefrac{2}{3}$\tabularnewline
\hline 
$\mathcal{B}'_{1}$ & $\nicefrac{1}{25}$ & $\nicefrac{1}{25}$ & $\nicefrac{1}{9}$ & $\nicefrac{2}{9}$ & $\nicefrac{1}{3}$ & $\nicefrac{4}{9}$ & $\nicefrac{5}{9}$ & $\nicefrac{2}{3}$\tabularnewline
$\mathcal{B}'_{2}$ & $\nicefrac{1}{5}$ & $\nicefrac{1}{5}$ & $\nicefrac{1}{5}$ & $\nicefrac{2}{9}$ & $\nicefrac{1}{3}$ & $\nicefrac{4}{9}$ & $\nicefrac{5}{9}$ & $\nicefrac{2}{3}$\tabularnewline
$\mathcal{B}'_{3}$ & $\nicefrac{9}{25}$ & $\nicefrac{9}{25}$ & $\nicefrac{9}{25}$ & $\nicefrac{9}{25}$ & $\nicefrac{9}{25}$ & $\nicefrac{4}{9}$ & $\nicefrac{5}{9}$ & $\nicefrac{2}{3}$\tabularnewline
\hline 
\end{tabular}
\par\end{centering}
\caption{\label{tab:A-table-showing-1}Contextuality value $d_{3}$ of disturbed
concatenated systems for the measures $\mathsf{CNT_{2}}$, $\mathsf{CNT_{3}}$,
and $\mathsf{CNTF}.$ The rest as in Table \ref{tab:A-table-showing}.}
\end{table}

The results presented here are just a fraction of the systems we explored
for this work: $6\times49$ undisturbed $\mathcal{A}$-$\mathcal{B}$
pairs and $125\times49$ disturbed $\mathcal{A}$-$\mathcal{B}$ pairs
(with many different subsystems producing identical profiles). The
contextuality curves we had to leave out in order not to clutter the
graphs and tables or multiply their number conform to the same pattern:
$\mathsf{CNT_{2}}$ is always additive, and the measures $\mathsf{CNT_{3}}$
and $\mathsf{CNTF}$ are subadditive because they conform to the rule
of maximum. 

With these regularities being established as inductive generalizations,
we can look for their analytic justification. Although this is not
essential for this paper, whose purpose is to introduce and demonstrate
the usefulness of the concept of a contextuality profile for discovery
of regularities, we outline these analytic arguments below.

\section{\label{sec:Outlines}Outlines of the proofs}

For $\mathsf{CNT_{2}}$, since $\mathcal{B}$ is noncontextual at
level 2, the value of $d_{2}$ is the $L_{1}$-distance between the
system $\mathcal{A}$ and the level-2 noncontextuality polytope. The
system and the polytope are defined in the space spanned by the axes
representing all pairwise probabilities. The value of $\Delta_{3}$
is the $L_{1}$-distance between the system $\mathcal{B}$ and the
level-3 noncontextuality polytope. Because $\mathcal{B}$ is noncontextual
at level 2, this distance is entirely within the space spanned by
the axes representing all triple probabilities. In the system $\mathcal{A*\mathcal{B}}$
the axes of these two spaces, of the pairwise and of the triple probabilities,
are combined as mutually orthogonal subspaces. By the nature of $L_{1}$,
therefore, the overall distance in this combined space is the sum
of the two subspace distances. If instead of the $L_{1}$-distance
we chose an $L_{p}$-distance with $p>1$, the overall distance would
have satisfied
\begin{equation}
d_{3}^{p}=d_{2}^{p}+\Delta_{3}^{p}.
\end{equation}

The argument establishing the rule of maximum is essentially the same
for $\mathsf{CNT_{3}}$ and $\mathsf{CNTF}$. Let us present the details
for the latter. 

With reference to (\ref{eq:CNTF inequality}) and (\ref{eq:CNTF measure}),
let $\mathbf{M}_{A}$ be the Boolean incidence matrix for system $\mathcal{A}$,
let $\mathbf{x}_{A}=\left(\alpha_{1},\ldots,\alpha_{K}\right)$ be
a vector of probabilities assigned to the $K$ combinations of values
of the variables in $\mathcal{A}$, and let $\mathbf{v}_{A}$ be the
vector of probabilities. We define analogously $\mathbf{M}_{B}$,
$\mathbf{x}_{B}=\left(\beta_{1},\ldots,\beta_{L}\right)$, and $\mathbf{v}_{B}$
for system $\mathcal{B}$, and $\mathbf{M}_{AB}$, $\mathbf{x}_{AB}=\left(\gamma_{11},\ldots,\gamma_{KL}\right)$,
and $\mathbf{v}_{AB}$ for system $\mathcal{A}*\mathcal{B}$. Here,
$\gamma_{ij}$ is assigned to the concatenation of the $i$th combination
of values in $\mathcal{A}$ and the $j$th combination of values in
$\mathcal{B}$; and
\begin{equation}
\mathbf{v}_{AB}=\left(\begin{array}{c}
\mathbf{v}_{A}\\
\mathbf{v}_{B}
\end{array}\right).\label{eq:v_AV}
\end{equation}
In matrix $\mathbf{M}_{A}$, let $a_{is}$ $\left(i=1,\ldots.K\right)$
be the entries of the row corresponding to the $s$th element $v_{A}^{\left(s\right)}$
of $\mathbf{v}_{A}$, and let $b_{jt}$ $\left(j=1,\ldots.L\right)$
be the entries of the row corresponding to the $t$th element $v_{B}^{\left(t\right)}$
of $\mathbf{v}_{B}$. For matrix $\mathbf{M}_{AB}$, let $c_{ijs}$
and $c_{ijt}$ be the entries of the rows corresponding, respectively,
to $v_{A}^{\left(s\right)}$ and to $v_{B}^{\left(t\right)}$ in (\ref{eq:v_AV}).
We have
\begin{equation}
\mathbf{M}_{A}\mathbf{x}_{A}=\left\{ \begin{array}{c}
\vdots\\
\sum_{i=1}^{K}a_{is}\alpha_{i}\\
\vdots
\end{array}\right.\leq\left\{ \begin{array}{c}
\vdots\\
v_{A}^{\left(s\right)}\\
\vdots
\end{array}\right.=\mathbf{v}_{A},
\end{equation}
\begin{equation}
\mathbf{M}_{B}\mathbf{x}_{B}=\left\{ \begin{array}{c}
\vdots\\
\sum_{j=1}^{L}b_{jt}\beta_{j}\\
\vdots
\end{array}\right.\leq\left\{ \begin{array}{c}
\vdots\\
v_{B}^{\left(t\right)}\\
\vdots
\end{array}\right.=\mathbf{v}_{B}\label{eq:ineqB}
\end{equation}
and
\begin{equation}
\mathbf{M}_{AB}\mathbf{x}_{AB}=\left\{ \begin{array}{c}
\vdots\\
\sum_{i=1}^{K}\sum_{j=1}^{L}c_{ijs}\gamma_{ij}\\
\vdots\\
\sum_{j=1}^{L}\sum_{i=1}^{K}c_{ijt}\gamma_{ij}\\
\vdots
\end{array}\right.\leq\left\{ \begin{array}{c}
\vdots\\
v_{A}^{\left(s\right)}\\
\vdots\\
v_{B}^{\left(t\right)}\\
\vdots
\end{array}\right.=\left(\begin{array}{c}
\mathbf{v}_{A}\\
\mathbf{v}_{B}
\end{array}\right).
\end{equation}
 Due to the structure of a concatenated system,
\begin{equation}
c_{ijs}=a_{is}
\end{equation}
irrespective of $j=1,\ldots.L$, and
\begin{equation}
c_{ijt}=b_{jt}
\end{equation}
irrespective of $i=1,\ldots.K$. Therefore the inequality for system
$\mathcal{A}*\mathcal{B}$ can be written as 
\begin{equation}
\mathbf{M}_{AB}\mathbf{x}_{AB}=\left\{ \begin{array}{c}
\vdots\\
\sum_{i=1}^{K}a_{is}\sum_{j=1}^{L}\gamma_{ij}\\
\vdots\\
\sum_{j=1}^{L}b_{jt}\sum_{i=1}^{K}\gamma_{ij}\\
\vdots
\end{array}\right.\leq\left\{ \begin{array}{c}
\vdots\\
v_{A}^{\left(s\right)}\\
\vdots\\
v_{B}^{\left(t\right)}\\
\vdots
\end{array}\right.=\left(\begin{array}{c}
\mathbf{v}_{A}\\
\mathbf{v}_{B}
\end{array}\right).\label{eq:ineqAB}
\end{equation}

Let us show now that

\begin{equation}
\left(\mathbf{1}^{\intercal}\mathbf{x}_{AB}\right)_{\max}\leq\min\left(\left(\mathbf{1}^{\intercal}\mathbf{x}_{A}\right)_{\max},\left(\mathbf{1}^{\intercal}\mathbf{x}_{B}\right)_{\max}\right).\label{eq:CNTF<=00003D}
\end{equation}
Indeed, if we had, e.g., $\left(\mathbf{1}^{\intercal}\mathbf{x}_{AB}\right)_{\max}>\left(\mathbf{1}^{\intercal}\mathbf{x}_{B}\right)_{\max}$,
then we could redefine the values of $\mathbf{x}_{B}$ as
\begin{equation}
\beta_{j}=\sum_{i=1}^{K}\gamma_{ij},j=1,\ldots,L,\label{eq:redefine beta}
\end{equation}
and, by substituting in (\ref{eq:ineqB}), obtain a coupling of $\mathcal{B}$
with a greater value of $\mathbf{1}^{\intercal}\mathbf{x}_{B}$ than
$\left(\mathbf{1}^{\intercal}\mathbf{x}_{B}\right)_{\max}$. The inequality
(\ref{eq:ineqB}) with the redefined vector will be preserved because
in holds in (\ref{eq:ineqAB}). 

It is also true that 
\begin{equation}
\left(\mathbf{1}^{\intercal}\mathbf{x}_{AB}\right)_{\max}\geq\min\left(\left(\mathbf{1}^{\intercal}\mathbf{x}_{A}\right)_{\max},\left(\mathbf{1}^{\intercal}\mathbf{x}_{B}\right)_{\max}\right),\label{eq:CNTF>=00003D}
\end{equation}
because if we had, e.g., $\left(\mathbf{1}^{\intercal}\mathbf{x}_{AB}\right)_{\max}<\left(\mathbf{1}^{\intercal}\mathbf{x}_{B}\right)_{\max}\leq\left(\mathbf{1}^{\intercal}\mathbf{x}_{A}\right)_{\max}$,
then we could redefine the values of $\mathbf{x}_{AB}$ as
\begin{equation}
\gamma_{ij}=\left\{ \begin{array}{ccc}
\beta_{j} & \textnormal{if} & i=1,\\
0 & \textnormal{if} & i>1,
\end{array}\right.\label{eq:redefine gamma}
\end{equation}
and obtain thereby a coupling of $\mathcal{A}*\mathcal{B}$ with a
greater value of $\mathbf{1}^{\intercal}\mathbf{x}_{AB}$ than $\left(\mathbf{1}^{\intercal}\mathbf{x}_{B}\right)_{\max}$.
The inequality in (\ref{eq:ineqAB}) with the redefined vector will
be preserved because it holds in (\ref{eq:ineqB}). The conjunction
of (\ref{eq:CNTF<=00003D}) and (\ref{eq:CNTF>=00003D}) yields the
rule of maximum, (\ref{eq:maximum rule}), because 
\[
\mathsf{CNTF}=1-\left(\mathbf{1}^{\intercal}\mathbf{x}\right)_{\max}.
\]

For $\mathsf{CNT_{3}}$, with reference to (\ref{eq:CNT3 equality})
and (\ref{eq:CNT3 measure}), we show that

\begin{equation}
\left(\mathbf{1}^{\intercal}\left|\mathbf{x}_{AB}\right|\right)_{\max}\geq\min\left(\left(\mathbf{1}^{\intercal}\left|\mathbf{x}_{A}\right|\right)_{\max},\left(\mathbf{1}^{\intercal}\left|\mathbf{x}_{B}\right|\right)_{\max}\right),\label{eq:CNTF>=00003D-1}
\end{equation}
because if, e.g., $\left(\mathbf{1}^{\intercal}\left|\mathbf{x}_{AB}\right|\right)_{\max}<\left(\mathbf{1}^{\intercal}\left|\mathbf{x}_{B}\right|\right)_{\max}$,
one could redefine the values of $\mathbf{x}_{AB}$ as in (\ref{eq:redefine gamma}),
and achieve an increase in $\mathbf{1}^{\intercal}\left|\mathbf{x}_{AB}\right|$.
By the same argument as above, the equality $\mathbf{M}_{AB}\mathbf{x}_{AB}=\mathbf{v}_{AB}$
will be preserved because $\mathbf{M}_{B}\mathbf{x}_{B}=\mathbf{v}_{B}$.

Also,
\begin{equation}
\left(\mathbf{1}^{\intercal}\left|\mathbf{x}_{AB}\right|\right)_{\max}\leq\min\left(\left(\mathbf{1}^{\intercal}\left|\mathbf{x}_{A}\right|\right)_{\max},\left(\mathbf{1}^{\intercal}\left|\mathbf{x}_{B}\right|\right)_{\max}\right),\label{eq:CNTF>=00003D-1-1}
\end{equation}
because if, e.g., $\left(\mathbf{1}^{\intercal}\left|\mathbf{x}_{AB}\right|\right)_{\max}>\left(\mathbf{1}^{\intercal}\left|\mathbf{x}_{B}\right|\right)_{\max}\geq\left(\mathbf{1}^{\intercal}\left|\mathbf{x}_{A}\right|\right)_{\max}$,
one could redefine the values of $\mathbf{x}_{B}$ as in (\ref{eq:redefine beta}),
and achieve an increase in $\mathbf{1}^{\intercal}\left|\mathbf{x}_{B}\right|$.
The equality $\mathbf{M}_{B}\mathbf{x}_{B}=\mathbf{v}_{B}$ will be
presented because $\mathbf{M}_{AB}\mathbf{x}_{AB}=\mathbf{v}_{AB}$.
(Note that the dimensions and entries of the vectors and matrices
are different for $\mathsf{CNT_{3}}$ and $\mathsf{CNTF}$, see Ref.
\citep{KujDzhMeasures})

\section{\label{sec:Other-systems}Hypercyclic systems}

Hypercyclic systems were introduced in Ref. \citep{hyper2-23} as
a set of systems that are both highly structured and sufficiently
diverse to form testing grounds for contextuality research. The subsystem
$\mathcal{A}$ in our concatenated systems is a cyclic system of rank
3 (a special case of a hypercyclic system), and the subsystem $\mathcal{B}$
was a hypercyclic system of order 3 and rank 4 but without a last
row (so abridged to speed up execution of the linear programs). The
order of a hypercyclic system is the number of variables in each context;
the rank is the number of the system's contexts (which is the same
as the number of questions). The variables in each row are cyclically
shifted clockwise with respect to the previous row. 

\begin{figure}
\begin{centering}
\includegraphics[scale=0.35]{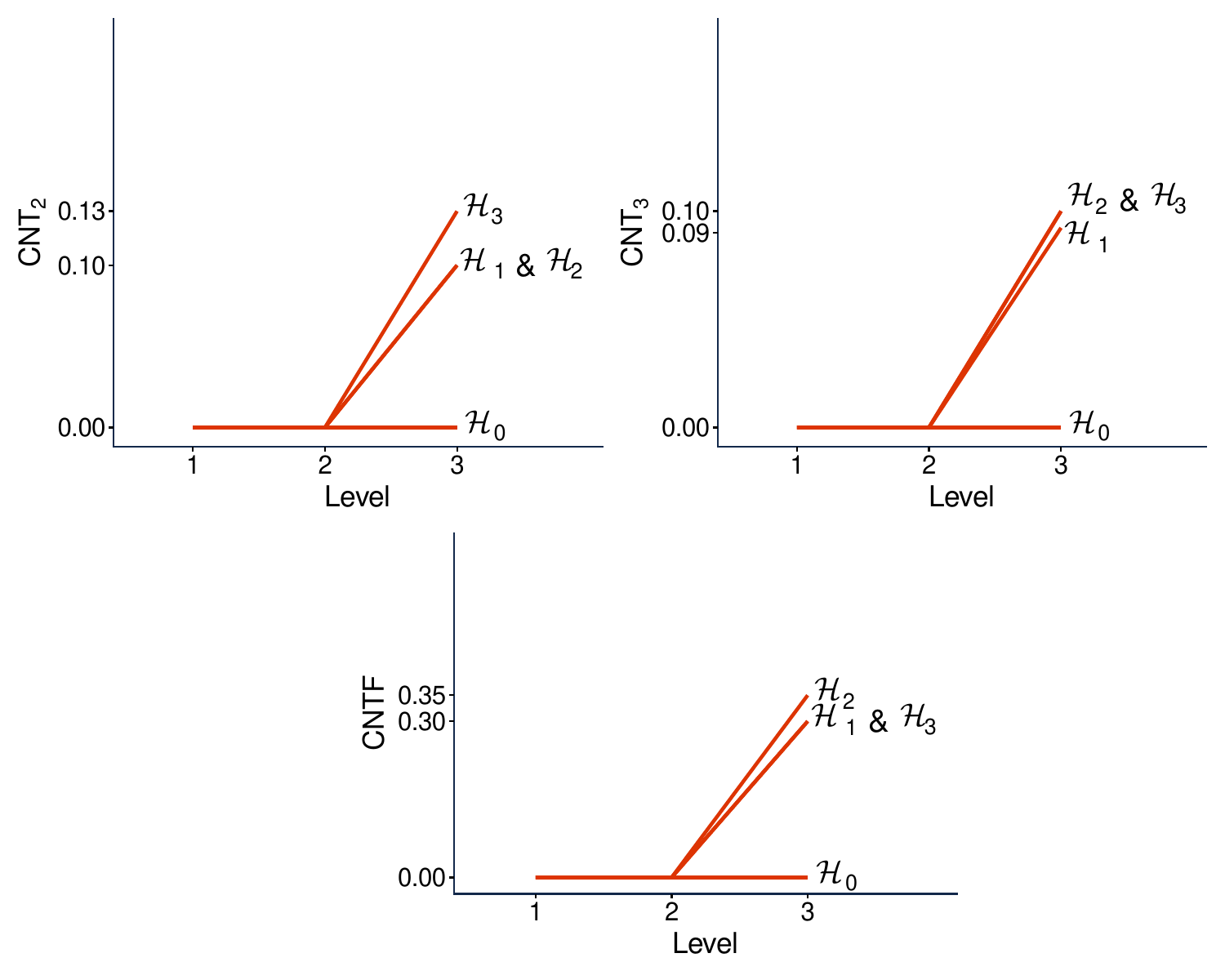}
\par\end{centering}
\caption{\label{fig:hypercyclic ud}Contextuality profiles for a selection
of undisturbed hypercyclic systems of order 3 and rank 4. }
\end{figure}

\begin{figure}
\begin{centering}
\includegraphics[scale=0.35]{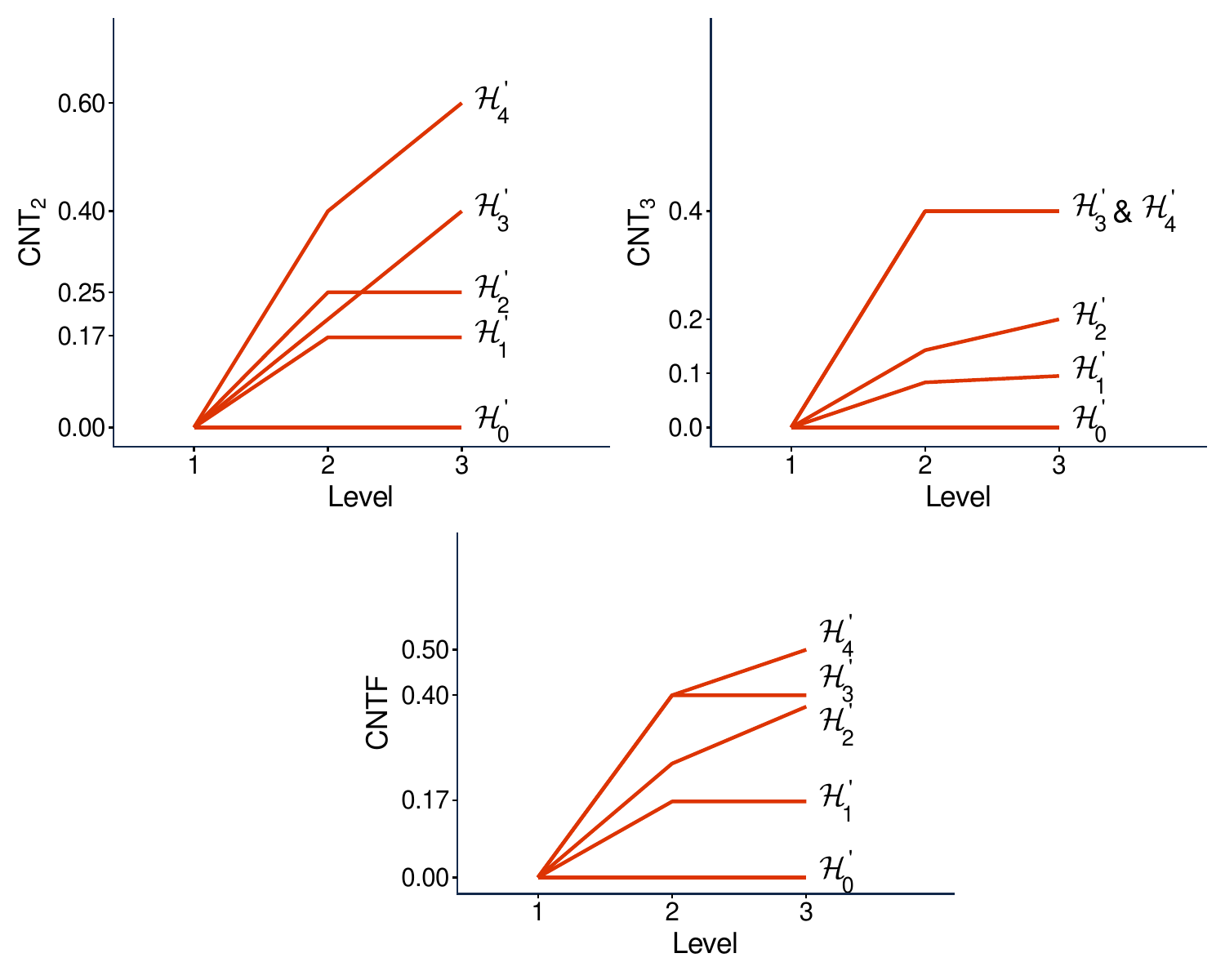}
\par\end{centering}
\caption{\label{fig:hypercyclic d}Contextuality profiles for a selection of
disturbed hypercyclic systems of order 3 and rank 4.}
\end{figure}

Figures \ref{fig:hypercyclic ud} and \ref{fig:hypercyclic d} present
contextuality profiles for a selection of hypercyclic systems of order
3 and rank 4:
\begin{equation}
\begin{array}{|c|c|c|c||c|}
\hline R_{1}^{1} & R_{2}^{1} & R_{3}^{1} &  & c=1\\
\hline  & R_{2}^{2} & R_{3}^{2} & R_{4}^{2} & 2\\
\hline R_{1}^{3} &  & R_{3}^{3} & R_{4}^{3} & 3\\
\hline R_{1}^{4} & R_{2}^{4} &  & R_{4}^{4} & 4\\
\hline\hline q=1 & 2 & 3 & 4 & 
\\\hline \end{array}\:.
\end{equation}
Figure \ref{fig:hypercyclic ud} has the property common to all undisturbed
hypercyclic systems: they are noncontextual at all but the final level
(in our case, level 3). Some disturbed hypercyclic systems have this
property too, but it does not hold generally.

\begin{figure}
\begin{centering}
\includegraphics[scale=0.35]{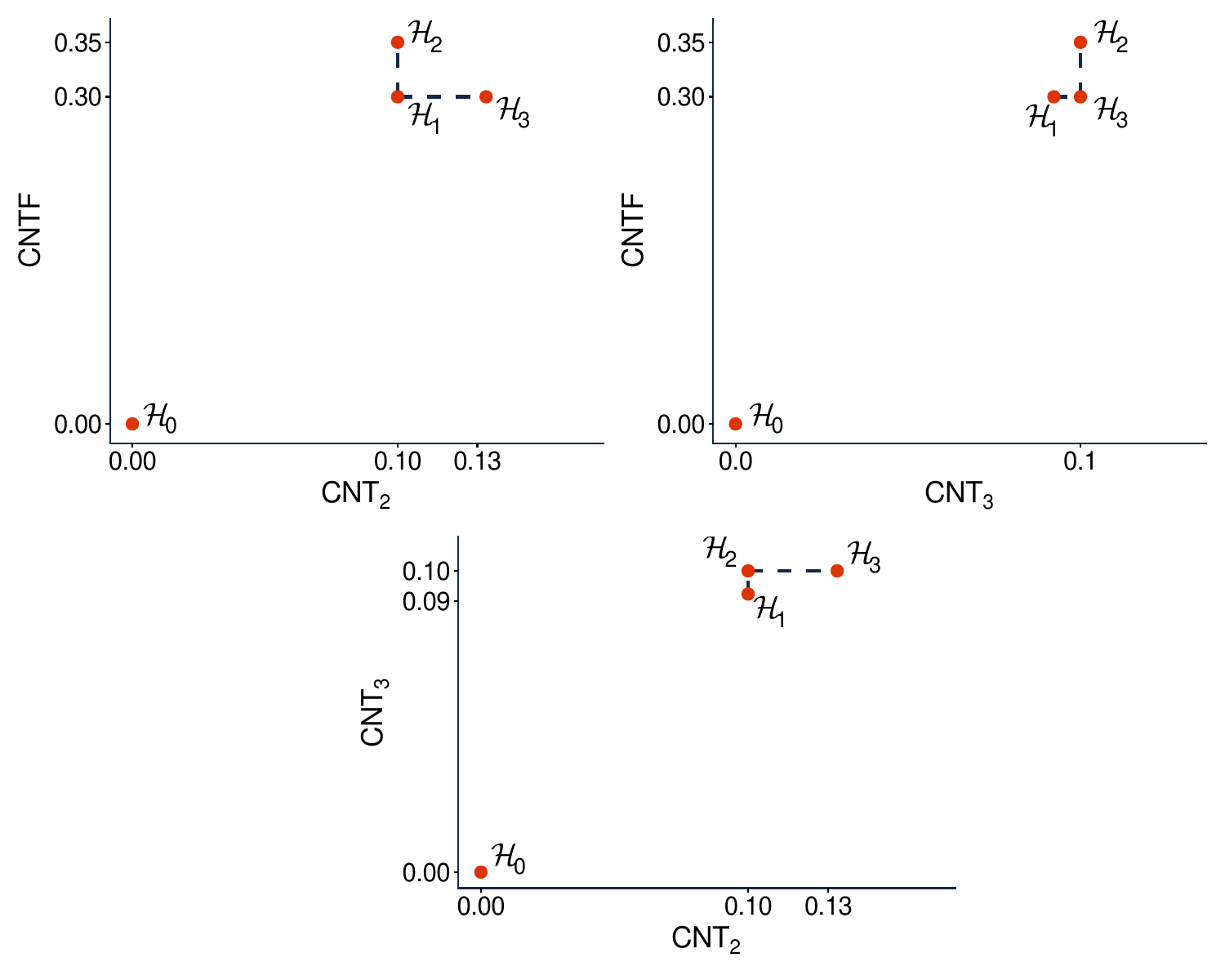}
\par\end{centering}
\caption{\label{fig:Contextuality-profiles-for-odd-2-2-1}Relationship between
the overall contextuality degrees generated by two contextuality measures
applied to our selection of undisturbed hypercyclic systems. The horizontal
lines show that the abscissa measure cannot be a function of the ordinate
one; the vertical lines show that the ordinate measure cannot be a
function of the abscissa one.}
\end{figure}

However, this property is not why we consider the hypercyclic systems
here. In this paper they serve another purpose. We know from Ref.
\citep{hyper2-23} that none of the three measures of contextuality
we studied, $\mathsf{CNT_{2}}$, $\mathsf{CNT_{3}}$, and $\mathsf{CNTF}$,
is a function of any other of them when considered across different
systems. Specifically, it was shown that, for any ordered pair of
these contextuality measures, e.g., $\left(\mathsf{CNTF},\mathsf{CNT}_{2}\right)$,
one can find two hypercyclic systems such that the first measure changes
from one of them to another while the second measure remains constant.
For our selection of undisturbed hypercyclic systems this is shown
in Figure \ref{fig:Contextuality-profiles-for-odd-2-2-1}. The interpretation
of this observation is that, unlike, say, $\mathsf{CNTF}$ and $\log\left(\mathsf{CNTF}\right)$,
the three measures $\mathsf{CNT_{2}}$, $\mathsf{CNTF}$, and $\mathsf{CNT_{3}}$
reflect pairwise distinct aspects of contextuality.

\begin{figure}
\begin{centering}
\includegraphics[scale=0.35]{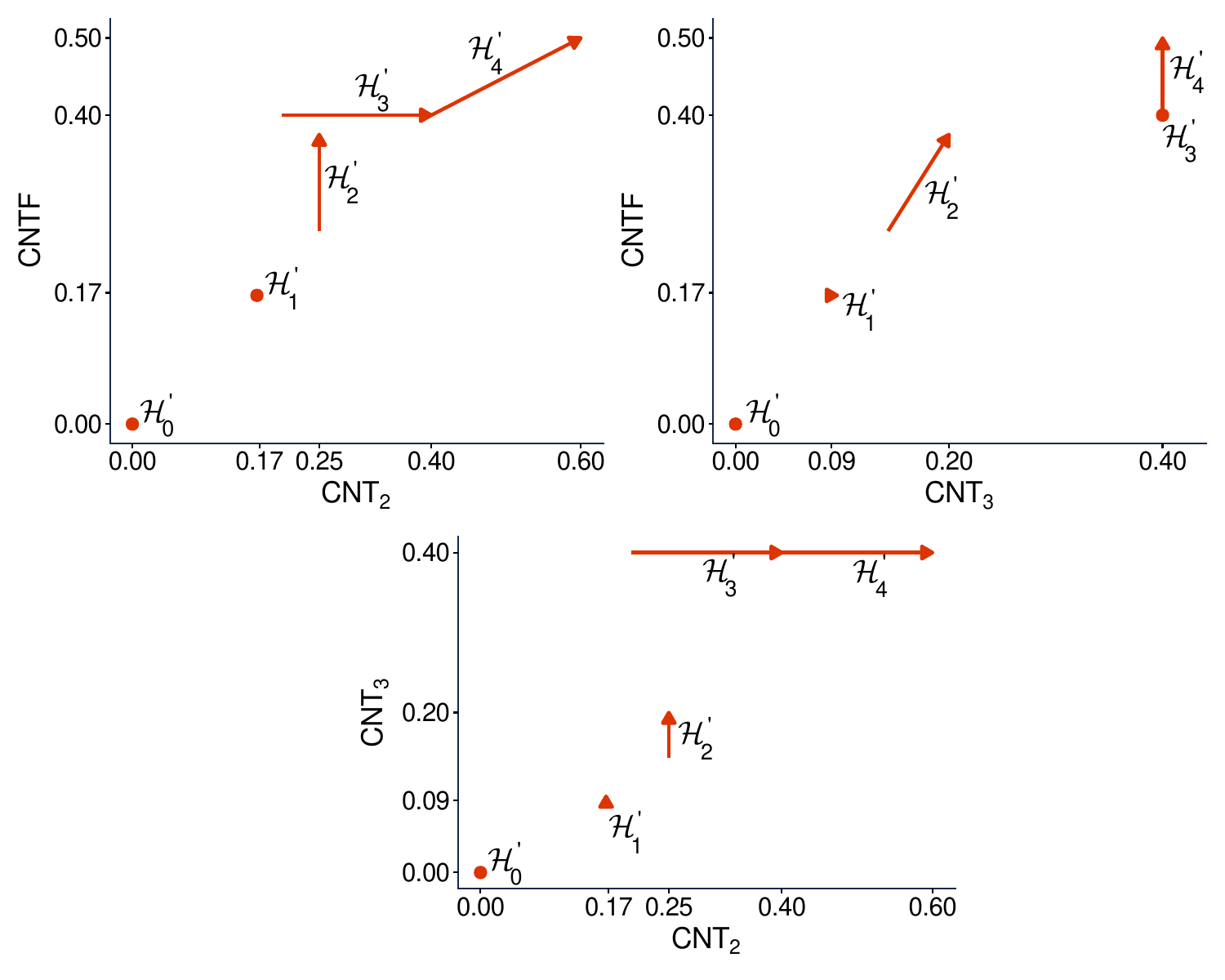}
\par\end{centering}
\caption{\label{fig:Contextuality-profiles-for-odd-2-2}Relationship between
the segments of contextuality profiles (between levels 2 and 3, as
indicated by arrows) generated by two measures applied to the same
disturbed hypercyclic system. The horizontal lines show that the abscissa
profile cannot be a function of the ordinate one; the vertical lines
show that the ordinate profile cannot be a function of the abscissa
one.}
\end{figure}

The question we pose in this paper is whether the same is true for
the contextual profiles generated by the three measures for one and
the same system. Figure \ref{fig:Contextuality-profiles-for-odd-2-2}
tells us that this is indeed the case: for any ordered pair of our
three measures one can find a system such that the first measure changes
between levels 2 and 3 while the second measure remains constant. 

\begin{figure}
\begin{centering}
\includegraphics[scale=0.35]{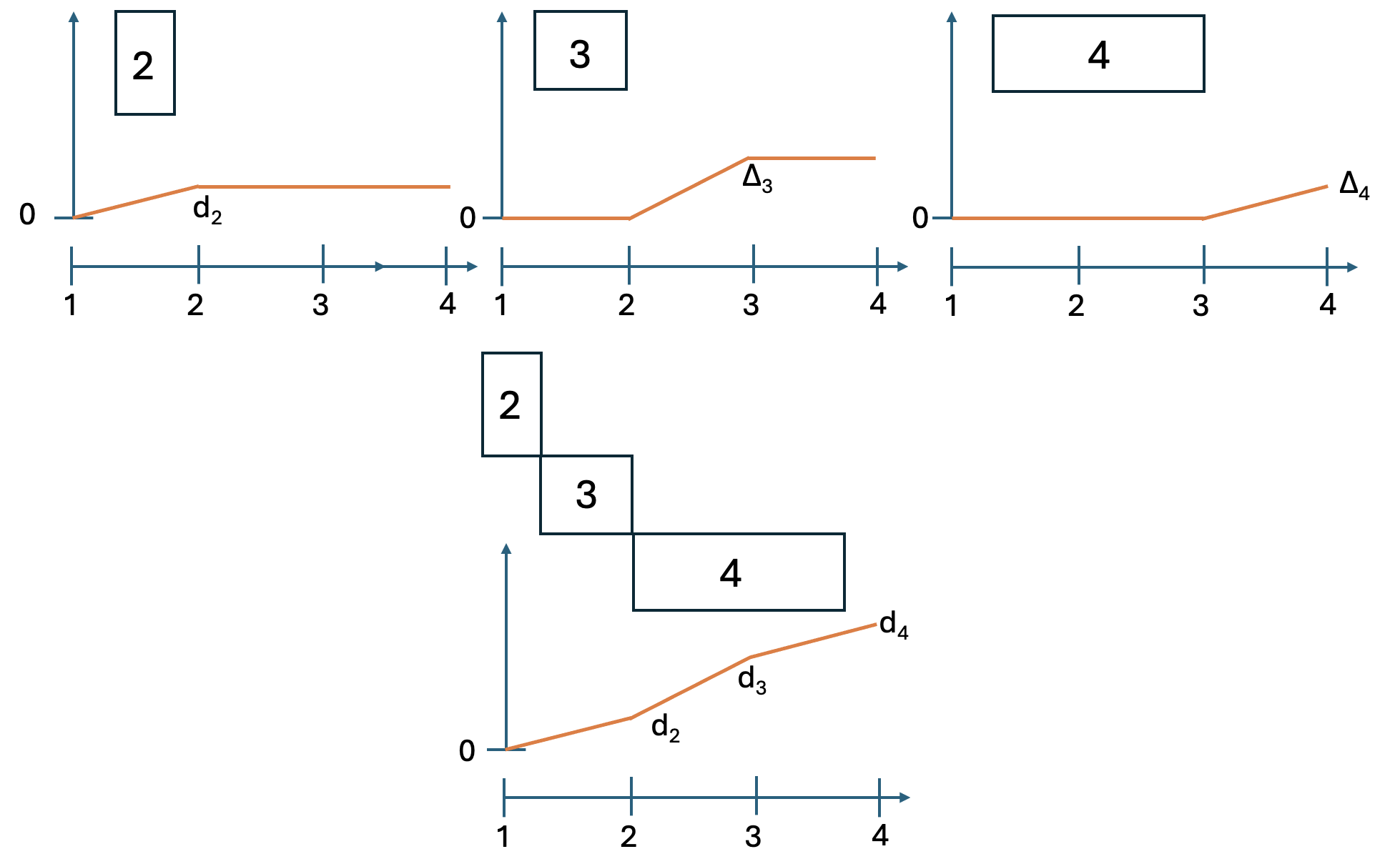}
\par\end{centering}
\caption{\label{fig:Contextuality-profiles-for-1}Contextuality profiles for
a triple-concatenation of systems. The boxes represent the systems
being concatenated, with the numbers in them indicating their final
level of contextuality.}
\end{figure}

\section{\label{sec:Conclusion}Concluding discussion}

We have introduced a new notion, that of a contextuality profile of
a system, and investigated some of its basic properties. We compared
the contextuality profiles of three well-constructed measures, $\mathsf{CNT_{2}}$,
$\mathsf{CNTF}$, and $\mathsf{CNT_{3}}$ using the method of concatenated
systems. We established that $\mathsf{CNT_{2}}$ profiles are additive
while $\mathsf{CNTF}$ and $\mathsf{CNT_{3}}$ profiles are subadditive,
because they conform to the rule of maximum. We have also established
that none of these three measures is a function of any other, not
only across different systems (which has been known previously), but
also within a system taken at different levels.

Note that concatenation can be used recursively, creating combinations
of three and more systems, as shown in Figure \ref{fig:Contextuality-profiles-for-1}.
For concatenations of $\mathcal{A}_{1},\ldots,\mathcal{A}_{k}$, the
arguments for the additivity of $\mathsf{CNT_{2}}$ and the maximum
rule for $\mathsf{CNTF}$ and $\mathsf{CNT_{3}}$ can be recursively
applied to show that 
\begin{equation}
\mathsf{CNT_{2}}=\sum_{i=2}^{k}\Delta_{i},\mathsf{CNT_{3}}=\max(\Delta'_{2},\ldots,\Delta'_{k}),\mathsf{CNTF}=\max(\Delta''_{2},\ldots,\Delta''_{k}),
\end{equation}
where we replaced $d_{2}$ with $\Delta_{2}$ for uniformity and added
primes to emphasize that the value of $\Delta_{i}$ is measure-specific.

This being only a concept paper, it leaves many questions unanswered.
From a mathematical point of view, much remains to be investigated
analytically regarding the properties of the contextuality profiles.
However, aside from the intrinsic mathematical interest, the main
substantive question is that of applicability: do the contextuality
profiles tell us something about other, independently defined properties
of the empirical entities described by the systems? Could, e.g., the
resource-theoretical aspects of the systems \citep{AbramBarbMans2017}
be better understood if we relate them to various aspects of contextuality
profiles rather than to the overall degree of contextuality only?
Another possible application was pointed out to us by Pawe{\l} Kurzy{\'n}ski
\citep{Pawel}: he suggested comparing the contextuality profiles
of the two well-known quantum systems, the Greenberger-Horne-Zeilinger
system (GHZ, \citep{GHZ1989}) and the W-state one \citep{WstateDuer2000}.
Both are known to be triple-entangled, but only the W-state system
is also pairwise-entangled. So it is of interest to see if this difference
is reflected in the contextuality profiles. It does not have to be,
because the relationship between quantum entanglement and contextuality
(including nonlocality as its special case) is not one-to-one. An
entangled system may very well be noncontextual, and it is only an
assumption based on an empirical generalization that any unentangled
system is noncontextual. 

The tripartite W-state is
\begin{equation}
\left|W\right\rangle =\frac{1}{\sqrt{3}}\LyXZeroWidthSpace\left(\left|001\right\rangle +\left|010\right\rangle +\left|100\right\rangle \right),\label{eq:W}
\end{equation}
 and the W-state system of random variables has the format

\begin{equation}
\begin{array}{|c|c|c|c|c|c||c|}
\hline A_{1}^{1} & A_{2}^{1} & A_{3}^{1} &  &  &  & c=1\\
\hline A_{1}^{2} &  &  &  & B_{2}^{2} & B_{3}^{2} & 2\\
\hline  & A_{2}^{3} &  & B_{1}^{3} &  & B_{3}^{3} & 3\\
\hline  &  & A_{3}^{4} & B_{1}^{4} & B_{2}^{4} &  & 4\\
\hline A_{1}^{5} & A_{2}^{5} &  &  &  & B_{3}^{5} & 5\\
\hline A_{1}^{6} &  & A_{3}^{6} &  & B_{2}^{6} &  & 6\\
\hline  & A_{2}^{7} & A_{3}^{7} & B_{1}^{7} &  &  & 7\\
\hline  &  &  & B_{1}^{8} & B_{2}^{8} & B_{3}^{8} & 8\\
\hline\hline q=Z_{1} & Z_{2} & Z_{3} & X_{1} & X_{2} & X_{3} & \mathcal{W}
\\\hline \end{array}\:,
\end{equation}
where $Z_{i}$ and $X_{i}$ denote the dichotomous measurements on
the $i$th particle along the $z$-axis and $x$-axis, respectively.
For convenience, we write $A_{i}^{c}$ and $B_{i}^{c}$ in place of
$R_{Z_{i}}^{c}$ and $R_{X_{i}}^{c}$. The distributions of the variables
in each context of $\mathcal{W}$, as derived from (\ref{eq:W}) are
shown in Appendix B. The contextuality of system $\mathcal{W}$ at
level 3 was established in Ref. \citep{Cabello2002}, and our computations
(using reduced couplings because the system is undisturbed) show that
at this level $\mathsf{CNT_{2}}=\nicefrac{1}{8}$, $\mathsf{CNT_{3}}=\nicefrac{1}{2}$
and $\mathsf{CNTF}=\nicefrac{1}{2}$. Ref. \citep{Turaetal2014} claims
to have established that the W-state system is also contextual at
level 2, aligning with the fact that its components are pairwise entangled.
However, this claim is not supported by our computations, showing
that $\mathcal{W}$ is noncontextual at level 2. Appendix B presents
the coupling of the system (Table \ref{tab:Distribution-of-a}) from
which all the pairwise probabilities in $\mathcal{W}$ (Table \ref{tab:Distribution-of-variables})
are obtained as marginals. This is a reduced coupling 
\begin{equation}
\left(S_{1},S_{2},S_{3},T_{1},T_{2},T_{3}\right)
\end{equation}
 such that
\begin{equation}
\left(S_{i},S_{j}\right)\deq\left(A_{i}^{c},A_{j}^{c}\right),\left(S_{i},T_{j}\right)\deq\left(A_{i}^{c},B_{j}^{c}\right),\left(T_{i},T_{j}\right)\deq\left(B_{i}^{c},B_{j}^{c}\right)
\end{equation}
for all $i,j\in$$\left\{ 1,2,3\right\} $ and any context $c$. 

The tripartite GHZ-state is
\begin{equation}
\left|GHZ\right\rangle =\frac{1}{\sqrt{2}}\LyXZeroWidthSpace\left(\LyXZeroWidthSpace\left|y_{+}\LyXZeroWidthSpace y_{+}\LyXZeroWidthSpace y_{+}\LyXZeroWidthSpace\right\rangle +\left|y_{-}\LyXZeroWidthSpace y_{-}\LyXZeroWidthSpace y_{-}\right\rangle \LyXZeroWidthSpace\right),\label{eq:GHZ}
\end{equation}
and the GHZ system of random variables has the same format as $\mathcal{W}$.
All its properties, however, can be established in its abridged version
\begin{equation}
\begin{array}{|c|c|c|c|c|c||c|}
\hline A_{1}^{1} & A_{2}^{1} & A_{3}^{1} &  &  &  & c=1\\
\hline A_{1}^{2} &  &  &  & B_{2}^{2} & B_{3}^{2} & 2\\
\hline  & A_{2}^{3} &  & B_{1}^{3} &  & B_{3}^{3} & 3\\
\hline  &  & A_{3}^{4} & B_{1}^{4} & B_{2}^{4} &  & 4\\
\hline\hline q=Z_{1} & Z_{2} & Z_{3} & X_{1} & X_{2} & X_{3} & \mathcal{GHZ}
\\\hline \end{array}\:,
\end{equation}
with the distributions shown in Appendix B. The contextuality of this
system on level 3 was also established in Ref. \citep{Cabello2002},
and our computations, using reduced couplings, yield $\mathsf{CNT_{2}}=\nicefrac{1}{4}$,
$\mathsf{CNT_{3}}=1$, $\mathsf{CNTF}=1$. The noncontextuality of
the GHZ system at level 2 is obvious, because the random variables
at this level are stochastically independent in every context (see
Table \ref{tab:Distribution-of-variables} in Appendix B). 

As we see, the contextuality profiles of the systems $\mathcal{GHZ}$
and $\mathcal{W}$ are qualitatively the same (0 at level 2 and some
positive value at level 3), in spite of the fact that a physical system
in the $\left|W\right\rangle $ state is pairwise entangled while
a physical system in the $\left|GHZ\right\rangle $ state is not.
More work is needed in search of more informative applications.\FloatBarrier

\section*{\label{sec:Appendix}Appendix A}

\setcounter{table}{0}
\renewcommand{\thetable}{A\arabic{table}}

Tables \ref{tab:Distributions-of-the}-\ref{tab:Distribution-of-the_hypercyclic-1-1}
show the distributions of all the systems mentioned in the main text.
Disturbed systems are indicated by primes ($\mathcal{A}'_{1}$, $\mathcal{B}'_{2}$,
$\mathcal{H}'_{0}$, etc.). 

\begin{table}[h]
\begin{adjustbox}{max width = \textwidth}\arraycolsep=2pt\tabcolsep=5pt%
\begin{tabular}{|c|c|c|c|c|c|c|}
\cline{2-7}
\multicolumn{1}{c|}{} & $\mathcal{A}_{0}$ & $\mathcal{A}_{1}$ & $\mathcal{A}_{2}$ & $\mathcal{A}_{3}$ & $\mathcal{A}_{4}$ & $\mathcal{A}_{5}$\tabularnewline
\hline 
$\begin{array}{c}
\begin{array}{c|c}
R_{1}^{1} & R_{2}^{1}\\\hline \end{array}\\
R_{1}^{1}=R_{2}^{1}
\end{array}$ & $\begin{array}{c}
\begin{array}{c|c}
0.5 & 0.5\\\hline \end{array}\\
0
\end{array}$ & $\begin{array}{c}
\begin{array}{c|c}
0.5 & 0.5\\\hline \end{array}\\
0
\end{array}$ & $\begin{array}{c}
\begin{array}{c|c}
0.5 & 0.5\\\hline \end{array}\\
0
\end{array}$ & $\begin{array}{c}
\begin{array}{c|c}
0.5 & 0.5\\\hline \end{array}\\
0
\end{array}$ & $\begin{array}{c}
\begin{array}{c|c}
0.5 & 0.5\\\hline \end{array}\\
0
\end{array}$ & $\begin{array}{c}
\begin{array}{c|c}
0.5 & 0.5\\\hline \end{array}\\
0
\end{array}$\tabularnewline
\hline 
$\begin{array}{c}
\begin{array}{c|c}
R_{2}^{2} & R_{3}^{2}\\\hline \end{array}\\
R_{2}^{2}=R_{3}^{2}
\end{array}$ & $\begin{array}{c}
\begin{array}{c|c}
0.5 & 0.5\\\hline \end{array}\\
0
\end{array}$ & $\begin{array}{c}
\begin{array}{c|c}
0.5 & 0.5\\\hline \end{array}\\
0
\end{array}$ & $\begin{array}{c}
\begin{array}{c|c}
0.5 & 0.5\\\hline \end{array}\\
0
\end{array}$ & $\begin{array}{c}
\begin{array}{c|c}
0.5 & 0.5\\\hline \end{array}\\
0
\end{array}$ & $\begin{array}{c}
\begin{array}{c|c}
0.5 & 0.5\\\hline \end{array}\\
0
\end{array}$ & $\begin{array}{c}
\begin{array}{c|c}
0.5 & 0.5\\\hline \end{array}\\
0
\end{array}$\tabularnewline
\hline 
$\begin{array}{c}
\begin{array}{c|c}
R_{3}^{3} & R_{4}^{3}\\\hline \end{array}\\
R_{3}^{3}=R_{4}^{3}
\end{array}$ & $\begin{array}{c}
\begin{array}{c|c}
0.5 & 0.5\\\hline \end{array}\\
0.5
\end{array}$ & $\begin{array}{c}
\begin{array}{c|c}
0.5 & 0.5\\\hline \end{array}\\
0.4
\end{array}$ & $\begin{array}{c}
\begin{array}{c|c}
0.5 & 0.5\\\hline \end{array}\\
0.3
\end{array}$ & $\begin{array}{c}
\begin{array}{c|c}
0.5 & 0.5\\\hline \end{array}\\
0.2
\end{array}$ & $\begin{array}{c}
\begin{array}{c|c}
0.5 & 0.5\\\hline \end{array}\\
0.1
\end{array}$ & $\begin{array}{c}
\begin{array}{c|c}
0.5 & 0.5\\\hline \end{array}\\
0
\end{array}$\tabularnewline
\hline 
\end{tabular}

\end{adjustbox}

\caption{\label{tab:Distributions-of-the}Distributions of the variables in
the selection of the undisturbed $\mathcal{A}$-subsystems used. All
variables are dichotomous ($\pm1$). Each number shows the probability
with which the corresponding variable(s) equal 1. Thus, $\tfrac{a|b}{c}$
in the second row means that $\Prob\left[R_{2}^{2}=1\right]=a$, $\Prob\left[R_{3}^{2}=1\right]=b$,
and $\Prob\left[R_{2}^{2}=R_{3}^{2}=1\right]=c$.}
\end{table}

\begin{table}[h]
\begin{adjustbox}{max width = \textwidth}\arraycolsep=2pt\tabcolsep=5pt%
\begin{tabular}{|c|c|c|c|c|c|c|c|}
\cline{2-8}
\multicolumn{1}{c|}{} & $\mathcal{A}'_{0}$ & $\mathcal{A}'_{1}$ & $\mathcal{A}'_{2}$ & $\mathcal{A}'_{3}$ & $\mathcal{A}'_{4}$ & $\mathcal{A}'_{5}$ & $\mathcal{A}'_{6}$\tabularnewline
\hline 
$\begin{array}{c}
\begin{array}{c|c}
R_{1}^{1} & R_{2}^{1}\\\hline \end{array}\\
R_{1}^{1}=R_{2}^{1}
\end{array}$ & $\begin{array}{c}
\begin{array}{c|c}
\nicefrac{4}{9} & \nicefrac{5}{9}\\\hline \end{array}\\
0
\end{array}$ & $\begin{array}{c}
\begin{array}{c|c}
\nicefrac{4}{9} & \nicefrac{5}{9}\\\hline \end{array}\\
0
\end{array}$ & $\begin{array}{c}
\begin{array}{c|c}
\nicefrac{4}{9} & \nicefrac{5}{9}\\\hline \end{array}\\
0
\end{array}$ & $\begin{array}{c}
\begin{array}{c|c}
\nicefrac{4}{9} & \nicefrac{5}{9}\\\hline \end{array}\\
0
\end{array}$ & $\begin{array}{c}
\begin{array}{c|c}
\nicefrac{4}{9} & \nicefrac{5}{9}\\\hline \end{array}\\
0
\end{array}$ & $\begin{array}{c}
\begin{array}{c|c}
\nicefrac{4}{9} & \nicefrac{5}{9}\\\hline \end{array}\\
0
\end{array}$ & $\begin{array}{c}
\begin{array}{c|c}
\nicefrac{4}{9} & \nicefrac{5}{9}\\\hline \end{array}\\
0
\end{array}$\tabularnewline
\hline 
$\begin{array}{c}
\begin{array}{c|c}
R_{2}^{2} & R_{3}^{2}\\\hline \end{array}\\
R_{2}^{2}=R_{3}^{2}
\end{array}$ & $\begin{array}{c}
\begin{array}{c|c}
\nicefrac{4}{9} & \nicefrac{5}{9}\\\hline \end{array}\\
\nicefrac{1}{9}
\end{array}$ & $\begin{array}{c}
\begin{array}{c|c}
\nicefrac{4}{9} & \nicefrac{5}{9}\\\hline \end{array}\\
\nicefrac{1}{18}
\end{array}$ & $\begin{array}{c}
\begin{array}{c|c}
\nicefrac{4}{9} & \nicefrac{5}{9}\\\hline \end{array}\\
0
\end{array}$ & $\begin{array}{c}
\begin{array}{c|c}
\nicefrac{4}{9} & \nicefrac{5}{9}\\\hline \end{array}\\
0
\end{array}$ & $\begin{array}{c}
\begin{array}{c|c}
\nicefrac{4}{9} & \nicefrac{5}{9}\\\hline \end{array}\\
0
\end{array}$ & $\begin{array}{c}
\begin{array}{c|c}
\nicefrac{4}{9} & \nicefrac{5}{9}\\\hline \end{array}\\
0
\end{array}$ & $\begin{array}{c}
\begin{array}{c|c}
\nicefrac{4}{9} & \nicefrac{5}{9}\\\hline \end{array}\\
0
\end{array}$\tabularnewline
\hline 
$\begin{array}{c}
\begin{array}{c|c}
R_{3}^{3} & R_{4}^{3}\\\hline \end{array}\\
R_{3}^{3}=R_{4}^{3}
\end{array}$ & $\begin{array}{c}
\begin{array}{c|c}
\nicefrac{4}{9} & \nicefrac{5}{9}\\\hline \end{array}\\
\nicefrac{2}{9}
\end{array}$ & $\begin{array}{c}
\begin{array}{c|c}
\nicefrac{4}{9} & \nicefrac{5}{9}\\\hline \end{array}\\
\nicefrac{2}{9}
\end{array}$ & $\begin{array}{c}
\begin{array}{c|c}
\nicefrac{4}{9} & \nicefrac{5}{9}\\\hline \end{array}\\
\nicefrac{2}{9}
\end{array}$ & $\begin{array}{c}
\begin{array}{c|c}
\nicefrac{4}{9} & \nicefrac{5}{9}\\\hline \end{array}\\
\nicefrac{3}{18}
\end{array}$ & $\begin{array}{c}
\begin{array}{c|c}
\nicefrac{4}{9} & \nicefrac{5}{9}\\\hline \end{array}\\
\nicefrac{1}{9}
\end{array}$ & $\begin{array}{c}
\begin{array}{c|c}
\nicefrac{4}{9} & \nicefrac{5}{9}\\\hline \end{array}\\
\nicefrac{1}{18}
\end{array}$ & $\begin{array}{c}
\begin{array}{c|c}
\nicefrac{4}{9} & \nicefrac{5}{9}\\\hline \end{array}\\
0
\end{array}$\tabularnewline
\hline 
\end{tabular}

\end{adjustbox}

\caption{\label{tab:The-same-as}The same as Table \ref{tab:Distributions-of-the}
but for the selection of the disturbed $\mathcal{A}$-subsystems used.}
\end{table}
\begin{table}
\begin{adjustbox}{max width = \textwidth}\arraycolsep=2pt\tabcolsep=5pt%
\begin{tabular}{|c|c|c|}
\cline{2-3}
\multicolumn{1}{c|}{} & $\mathcal{B}_{1}$ & $\mathcal{B}_{2}$\tabularnewline
\hline 
$\begin{array}{c}
\begin{array}{c|c|c}
R_{1}^{1} & R_{2}^{1} & R_{3}^{1}\\\hline \end{array}\\
\begin{array}{c|c|c}
R_{1}^{1}=R_{2}^{1} & R_{2}^{1}=R_{3}^{1} & R_{1}^{1}=R_{3}^{1}\\\hline \end{array}\\
R_{1}^{1}=R_{2}^{1}=R_{3}^{1}
\end{array}$ & $\begin{array}{c}
\begin{array}{c|c|c}
\nicefrac{1}{2} & \nicefrac{1}{2} & \nicefrac{1}{2}\\\hline \end{array}\\
\begin{array}{c|c|c}
\nicefrac{1}{4} & \nicefrac{1}{4} & \nicefrac{1}{4}\\\hline \end{array}\\
\nicefrac{1}{8}
\end{array}$ & $\begin{array}{c}
\begin{array}{c|c|c}
\nicefrac{1}{2} & \nicefrac{1}{2} & \nicefrac{1}{2}\\\hline \end{array}\\
\begin{array}{c|c|c}
\nicefrac{1}{4} & \nicefrac{1}{4} & \nicefrac{1}{4}\\\hline \end{array}\\
\nicefrac{1}{8}
\end{array}$\tabularnewline
\hline 
$\begin{array}{c}
\begin{array}{c|c|c}
R_{2}^{2} & R_{3}^{2} & R_{4}^{2}\\\hline \end{array}\\
\begin{array}{c|c|c}
R_{2}^{2}=R_{3}^{2} & R_{3}^{2}=R_{4}^{2} & R_{2}^{2}=R_{4}^{2}\\\hline \end{array}\\
R_{2}^{2}=R_{3}^{2}=R_{4}^{2}
\end{array}$ & $\begin{array}{c}
\begin{array}{c|c|c}
\nicefrac{1}{2} & \nicefrac{1}{2} & \nicefrac{1}{2}\\\hline \end{array}\\
\begin{array}{c|c|c}
\nicefrac{1}{4} & \nicefrac{1}{4} & \nicefrac{1}{4}\\\hline \end{array}\\
0
\end{array}$ & $\begin{array}{c}
\begin{array}{c|c|c}
\nicefrac{1}{2} & \nicefrac{1}{2} & \nicefrac{1}{2}\\\hline \end{array}\\
\begin{array}{c|c|c}
\nicefrac{1}{4} & \nicefrac{1}{4} & \nicefrac{1}{4}\\\hline \end{array}\\
\nicefrac{1}{6}
\end{array}$\tabularnewline
\hline 
$\begin{array}{c}
\begin{array}{c|c|c}
R_{3}^{3} & R_{4}^{3} & R_{1}^{3}\\\hline \end{array}\\
\begin{array}{c|c|c}
R_{3}^{3}=R_{4}^{3} & R_{4}^{3}=R_{1}^{3} & R_{3}^{3}=R_{1}^{3}\\\hline \end{array}\\
R_{3}^{3}=R_{4}^{3}=R_{1}^{3}
\end{array}$ & $\begin{array}{c}
\begin{array}{c|c|c}
\nicefrac{1}{2} & \nicefrac{1}{2} & \nicefrac{1}{2}\\\hline \end{array}\\
\begin{array}{c|c|c}
\nicefrac{1}{4} & \nicefrac{1}{4} & \nicefrac{1}{4}\\\hline \end{array}\\
\nicefrac{1}{4}
\end{array}$ & $\begin{array}{c}
\begin{array}{c|c|c}
\nicefrac{1}{2} & \nicefrac{1}{2} & \nicefrac{1}{2}\\\hline \end{array}\\
\begin{array}{c|c|c}
\nicefrac{1}{4} & \nicefrac{1}{4} & \nicefrac{1}{4}\\\hline \end{array}\\
0
\end{array}$\tabularnewline
\hline 
\end{tabular}

\end{adjustbox}

\caption{\label{tab:Distributions-of-the-1}Distributions of the variables
in the selection of the undisturbed $\mathcal{B}$-subsystems used.
As in Table \ref{tab:Distributions-of-the}, each number shows the
probability with which the corresponding variable(s) equal 1. Thus,
$\tfrac{\tfrac{a|b|c}{d|e|f}}{g}$ in the second row means that $\Prob\left[R_{2}^{2}=1\right]=a$,
$\Prob\left[R_{2}^{2}=R_{3}^{2}=1\right]=d$, $\Prob\left[R_{2}^{2}=R_{3}^{2}=R_{4}^{2}=1\right]=g$,
etc.}
\end{table}

\begin{table}
\begin{adjustbox}{max width = \textwidth}\arraycolsep=2pt\tabcolsep=5pt%
\begin{tabular}{|c|c|c|c|}
\cline{2-4}
\multicolumn{1}{c|}{} & $\mathcal{B}'_{1}$ & $\mathcal{B}'_{2}$ & $\mathcal{B}'_{3}$\tabularnewline
\hline 
$\begin{array}{c}
\begin{array}{c|c|c}
R_{1}^{1} & R_{2}^{1} & R_{3}^{1}\\\hline \end{array}\\
\begin{array}{c|c|c}
R_{1}^{1}=R_{2}^{1} & R_{2}^{1}=R_{3}^{1} & R_{1}^{1}=R_{3}^{1}\\\hline \end{array}\\
R_{1}^{1}=R_{2}^{1}=R_{3}^{1}
\end{array}$ & $\begin{array}{c}
\begin{array}{c|c|c}
\nicefrac{13}{25} & \nicefrac{1}{2} & \nicefrac{12}{25}\\\hline \end{array}\\
\begin{array}{c|c|c}
\nicefrac{13}{50} & \nicefrac{6}{25} & \nicefrac{6}{25}\\\hline \end{array}\\
\nicefrac{3}{25}
\end{array}$ & $\begin{array}{c}
\begin{array}{c|c|c}
\nicefrac{13}{25} & \nicefrac{1}{2} & \nicefrac{12}{25}\\\hline \end{array}\\
\begin{array}{c|c|c}
\nicefrac{13}{50} & \nicefrac{6}{25} & \nicefrac{6}{25}\\\hline \end{array}\\
\nicefrac{3}{25}
\end{array}$ & $\begin{array}{c}
\begin{array}{c|c|c}
\nicefrac{13}{25} & \nicefrac{1}{2} & \nicefrac{12}{25}\\\hline \end{array}\\
\begin{array}{c|c|c}
\nicefrac{13}{50} & \nicefrac{6}{25} & \nicefrac{6}{25}\\\hline \end{array}\\
\nicefrac{3}{25}
\end{array}$\tabularnewline
\hline 
$\begin{array}{c}
\begin{array}{c|c|c}
R_{2}^{2} & R_{3}^{2} & R_{4}^{2}\\\hline \end{array}\\
\begin{array}{c|c|c}
R_{2}^{2}=R_{3}^{2} & R_{3}^{2}=R_{4}^{2} & R_{2}^{2}=R_{4}^{2}\\\hline \end{array}\\
R_{2}^{2}=R_{3}^{2}=R_{4}^{2}
\end{array}$ & $\begin{array}{c}
\begin{array}{c|c|c}
\nicefrac{13}{25} & \nicefrac{1}{2} & \nicefrac{12}{25}\\\hline \end{array}\\
\begin{array}{c|c|c}
\nicefrac{13}{50} & \nicefrac{6}{25} & \nicefrac{6}{25}\\\hline \end{array}\\
0
\end{array}$ & $\begin{array}{c}
\begin{array}{c|c|c}
\nicefrac{13}{25} & \nicefrac{1}{2} & \nicefrac{12}{25}\\\hline \end{array}\\
\begin{array}{c|c|c}
\nicefrac{13}{50} & \nicefrac{6}{25} & \nicefrac{6}{25}\\\hline \end{array}\\
0
\end{array}$ & $\begin{array}{c}
\begin{array}{c|c|c}
\nicefrac{13}{25} & \nicefrac{1}{2} & \nicefrac{12}{25}\\\hline \end{array}\\
\begin{array}{c|c|c}
\nicefrac{13}{50} & \nicefrac{6}{25} & \nicefrac{6}{25}\\\hline \end{array}\\
0
\end{array}$\tabularnewline
\hline 
$\begin{array}{c}
\begin{array}{c|c|c}
R_{3}^{3} & R_{4}^{3} & R_{1}^{3}\\\hline \end{array}\\
\begin{array}{c|c|c}
R_{3}^{3}=R_{4}^{3} & R_{4}^{3}=R_{1}^{3} & R_{3}^{3}=R_{1}^{3}\\\hline \end{array}\\
R_{3}^{3}=R_{4}^{3}=R_{1}^{3}
\end{array}$ & $\begin{array}{c}
\begin{array}{c|c|c}
\nicefrac{13}{25} & \nicefrac{1}{2} & \nicefrac{12}{25}\\\hline \end{array}\\
\begin{array}{c|c|c}
\nicefrac{13}{50} & \nicefrac{6}{25} & \nicefrac{6}{25}\\\hline \end{array}\\
\nicefrac{2}{25}
\end{array}$ & $\begin{array}{c}
\begin{array}{c|c|c}
\nicefrac{13}{25} & \nicefrac{1}{2} & \nicefrac{12}{25}\\\hline \end{array}\\
\begin{array}{c|c|c}
\nicefrac{13}{50} & \nicefrac{6}{25} & \nicefrac{6}{25}\\\hline \end{array}\\
\nicefrac{1}{25}
\end{array}$ & $\begin{array}{c}
\begin{array}{c|c|c}
\nicefrac{13}{25} & \nicefrac{1}{2} & \nicefrac{12}{25}\\\hline \end{array}\\
\begin{array}{c|c|c}
\nicefrac{13}{50} & \nicefrac{6}{25} & \nicefrac{6}{25}\\\hline \end{array}\\
0
\end{array}$\tabularnewline
\hline 
\end{tabular}

\end{adjustbox}

\caption{\label{tab:The-same-as-1}The same as Table \ref{tab:Distributions-of-the-1},
but for the selection of the disturbed $\mathcal{B}$-subsystems used.}
\end{table}

\begin{table}
\begin{adjustbox}{max width = \textwidth}\arraycolsep=2pt\tabcolsep=5pt%
\begin{tabular}{|c|c|c|c|c|}
\cline{2-5}
\multicolumn{1}{c|}{} & $\mathcal{H}{}_{0}$ & $\mathcal{H}{}_{1}$ & $\mathcal{H}{}_{2}$ & $\mathcal{H}{}_{3}$\tabularnewline
\hline 
$\begin{array}{c}
\begin{array}{c|c|c}
R_{1}^{1} & R_{2}^{1} & R_{3}^{1}\\\hline \end{array}\\
\begin{array}{c|c|c}
R_{1}^{1}=R_{2}^{1} & R_{2}^{1}=R_{3}^{1} & R_{1}^{1}=R_{3}^{1}\\\hline \end{array}\\
R_{1}^{1}=R_{2}^{1}=R_{3}^{1}
\end{array}$ & $\begin{array}{c}
\begin{array}{c|c|c}
\nicefrac{1}{2} & \nicefrac{1}{2} & \nicefrac{1}{2}\\\hline \end{array}\\
\begin{array}{c|c|c}
\nicefrac{1}{4} & \nicefrac{1}{4} & \nicefrac{1}{4}\\\hline \end{array}\\
\nicefrac{1}{5}
\end{array}$ & $\begin{array}{c}
\begin{array}{c|c|c}
\nicefrac{1}{2} & \nicefrac{1}{2} & \nicefrac{1}{2}\\\hline \end{array}\\
\begin{array}{c|c|c}
\nicefrac{1}{4} & \nicefrac{1}{4} & \nicefrac{1}{4}\\\hline \end{array}\\
\nicefrac{1}{5}
\end{array}$ & $\begin{array}{c}
\begin{array}{c|c|c}
\nicefrac{1}{2} & \nicefrac{1}{2} & \nicefrac{1}{2}\\\hline \end{array}\\
\begin{array}{c|c|c}
\nicefrac{1}{4} & \nicefrac{1}{4} & \nicefrac{1}{4}\\\hline \end{array}\\
\nicefrac{1}{4}
\end{array}$ & $\begin{array}{c}
\begin{array}{c|c|c}
\nicefrac{1}{2} & \nicefrac{1}{2} & \nicefrac{1}{2}\\\hline \end{array}\\
\begin{array}{c|c|c}
\nicefrac{1}{4} & \nicefrac{1}{4} & \nicefrac{1}{4}\\\hline \end{array}\\
\nicefrac{1}{5}
\end{array}$\tabularnewline
\hline 
$\begin{array}{c}
\begin{array}{c|c|c}
R_{2}^{2} & R_{3}^{2} & R_{4}^{2}\\\hline \end{array}\\
\begin{array}{c|c|c}
R_{2}^{2}=R_{3}^{2} & R_{3}^{2}=R_{4}^{2} & R_{2}^{2}=R_{4}^{2}\\\hline \end{array}\\
R_{2}^{2}=R_{3}^{2}=R_{4}^{2}
\end{array}$ & $\begin{array}{c}
\begin{array}{c|c|c}
\nicefrac{1}{2} & \nicefrac{1}{2} & \nicefrac{1}{2}\\\hline \end{array}\\
\begin{array}{c|c|c}
\nicefrac{1}{4} & \nicefrac{1}{4} & \nicefrac{1}{4}\\\hline \end{array}\\
\nicefrac{3}{20}
\end{array}$ & $\begin{array}{c}
\begin{array}{c|c|c}
\nicefrac{1}{2} & \nicefrac{1}{2} & \nicefrac{1}{2}\\\hline \end{array}\\
\begin{array}{c|c|c}
\nicefrac{1}{4} & \nicefrac{1}{4} & \nicefrac{1}{4}\\\hline \end{array}\\
\nicefrac{1}{5}
\end{array}$ & $\begin{array}{c}
\begin{array}{c|c|c}
\nicefrac{1}{2} & \nicefrac{1}{2} & \nicefrac{1}{2}\\\hline \end{array}\\
\begin{array}{c|c|c}
\nicefrac{1}{4} & \nicefrac{1}{4} & \nicefrac{1}{4}\\\hline \end{array}\\
\nicefrac{1}{5}
\end{array}$ & $\begin{array}{c}
\begin{array}{c|c|c}
\nicefrac{1}{2} & \nicefrac{1}{2} & \nicefrac{1}{2}\\\hline \end{array}\\
\begin{array}{c|c|c}
\nicefrac{1}{4} & \nicefrac{1}{4} & \nicefrac{1}{4}\\\hline \end{array}\\
\nicefrac{1}{5}
\end{array}$\tabularnewline
\hline 
$\begin{array}{c}
\begin{array}{c|c|c}
R_{3}^{3} & R_{4}^{3} & R_{1}^{3}\\\hline \end{array}\\
\begin{array}{c|c|c}
R_{3}^{3}=R_{4}^{3} & R_{4}^{3}=R_{1}^{3} & R_{3}^{3}=R_{1}^{3}\\\hline \end{array}\\
R_{3}^{3}=R_{4}^{3}=R_{1}^{3}
\end{array}$ & $\begin{array}{c}
\begin{array}{c|c|c}
\nicefrac{1}{2} & \nicefrac{1}{2} & \nicefrac{1}{2}\\\hline \end{array}\\
\begin{array}{c|c|c}
\nicefrac{1}{4} & \nicefrac{1}{4} & \nicefrac{1}{4}\\\hline \end{array}\\
\nicefrac{3}{20}
\end{array}$ & $\begin{array}{c}
\begin{array}{c|c|c}
\nicefrac{1}{2} & \nicefrac{1}{2} & \nicefrac{1}{2}\\\hline \end{array}\\
\begin{array}{c|c|c}
\nicefrac{1}{4} & \nicefrac{1}{4} & \nicefrac{1}{4}\\\hline \end{array}\\
\nicefrac{1}{5}
\end{array}$ & $\begin{array}{c}
\begin{array}{c|c|c}
\nicefrac{1}{2} & \nicefrac{1}{2} & \nicefrac{1}{2}\\\hline \end{array}\\
\begin{array}{c|c|c}
\nicefrac{1}{4} & \nicefrac{1}{4} & \nicefrac{1}{4}\\\hline \end{array}\\
\nicefrac{3}{20}
\end{array}$ & $\begin{array}{c}
\begin{array}{c|c|c}
\nicefrac{1}{2} & \nicefrac{1}{2} & \nicefrac{1}{2}\\\hline \end{array}\\
\begin{array}{c|c|c}
\nicefrac{1}{4} & \nicefrac{1}{4} & \nicefrac{1}{4}\\\hline \end{array}\\
\nicefrac{1}{5}
\end{array}$\tabularnewline
\hline 
$\begin{array}{c}
\begin{array}{c|c|c}
R_{4}^{4} & R_{1}^{4} & R_{2}^{4}\\\hline \end{array}\\
\begin{array}{c|c|c}
R_{4}^{4}=R_{1}^{4} & R_{1}^{4}=R_{2}^{4} & R_{4}^{4}=R_{2}^{4}\\\hline \end{array}\\
R_{4}^{4}=R_{1}^{4}=R_{2}^{4}
\end{array}$ & $\begin{array}{c}
\begin{array}{c|c|c}
\nicefrac{1}{2} & \nicefrac{1}{2} & \nicefrac{1}{2}\\\hline \end{array}\\
\begin{array}{c|c|c}
\nicefrac{1}{4} & \nicefrac{1}{4} & \nicefrac{1}{4}\\\hline \end{array}\\
\nicefrac{3}{20}
\end{array}$ & $\begin{array}{c}
\begin{array}{c|c|c}
\nicefrac{1}{2} & \nicefrac{1}{2} & \nicefrac{1}{2}\\\hline \end{array}\\
\begin{array}{c|c|c}
\nicefrac{1}{4} & \nicefrac{1}{4} & \nicefrac{1}{4}\\\hline \end{array}\\
\nicefrac{3}{20}
\end{array}$ & $\begin{array}{c}
\begin{array}{c|c|c}
\nicefrac{1}{2} & \nicefrac{1}{2} & \nicefrac{1}{2}\\\hline \end{array}\\
\begin{array}{c|c|c}
\nicefrac{1}{4} & \nicefrac{1}{4} & \nicefrac{1}{4}\\\hline \end{array}\\
\nicefrac{3}{20}
\end{array}$ & $\begin{array}{c}
\begin{array}{c|c|c}
\nicefrac{1}{2} & \nicefrac{1}{2} & \nicefrac{1}{2}\\\hline \end{array}\\
\begin{array}{c|c|c}
\nicefrac{1}{4} & \nicefrac{1}{4} & \nicefrac{1}{4}\\\hline \end{array}\\
\nicefrac{1}{5}
\end{array}$\tabularnewline
\hline 
\end{tabular}

\end{adjustbox}

\caption{\label{tab:Distribution-of-the_hypercyclic-1}Distributions in the
selection of the undisturbed hypercyclic systems used. The format
is the same as in Table \ref{tab:Distributions-of-the-1}.}
\end{table}

\begin{table}
\begin{adjustbox}{max width = \textwidth}\arraycolsep=2pt\tabcolsep=5pt%
\begin{tabular}{|c|c|c|c|c|c|}
\cline{2-6}
\multicolumn{1}{c|}{} & $\mathcal{H}'_{0}$ & $\mathcal{H}'_{1}$ & $\mathcal{H}'_{2}$ & $\mathcal{H}'_{3}$ & $\mathcal{H}'_{4}$\tabularnewline
\hline 
$\begin{array}{c}
\begin{array}{c|c|c}
R_{1}^{1} & R_{2}^{1} & R_{3}^{1}\\\hline \end{array}\\
\begin{array}{c|c|c}
R_{1}^{1}=R_{2}^{1} & R_{2}^{1}=R_{3}^{1} & R_{1}^{1}=R_{3}^{1}\\\hline \end{array}\\
R_{1}^{1}=R_{2}^{1}=R_{3}^{1}
\end{array}$ & $\begin{array}{c}
\begin{array}{c|c|c}
\nicefrac{2}{3} & \nicefrac{2}{3} & \nicefrac{1}{2}\\\hline \end{array}\\
\begin{array}{c|c|c}
\nicefrac{1}{3} & \nicefrac{1}{3} & \nicefrac{1}{3}\\\hline \end{array}\\
\nicefrac{1}{6}
\end{array}$ & $\begin{array}{c}
\begin{array}{c|c|c}
\nicefrac{1}{3} & \nicefrac{1}{2} & \nicefrac{2}{3}\\\hline \end{array}\\
\begin{array}{c|c|c}
\nicefrac{1}{6} & \nicefrac{1}{3} & \nicefrac{1}{3}\\\hline \end{array}\\
\nicefrac{1}{6}
\end{array}$ & $\begin{array}{c}
\begin{array}{c|c|c}
\nicefrac{1}{4} & \nicefrac{1}{4} & \nicefrac{1}{4}\\\hline \end{array}\\
\begin{array}{c|c|c}
0 & 0 & 0\\\hline \end{array}\\
0
\end{array}$ & $\begin{array}{c}
\begin{array}{c|c|c}
\nicefrac{3}{5} & \nicefrac{3}{5} & \nicefrac{1}{5}\\\hline \end{array}\\
\begin{array}{c|c|c}
\nicefrac{2}{5} & \nicefrac{1}{5} & \nicefrac{1}{5}\\\hline \end{array}\\
\nicefrac{1}{5}
\end{array}$ & $\begin{array}{c}
\begin{array}{c|c|c}
\nicefrac{3}{5} & \nicefrac{3}{5} & \nicefrac{1}{5}\\\hline \end{array}\\
\begin{array}{c|c|c}
\nicefrac{2}{5} & \nicefrac{1}{5} & \nicefrac{1}{5}\\\hline \end{array}\\
\nicefrac{1}{5}
\end{array}$\tabularnewline
\hline 
$\begin{array}{c}
\begin{array}{c|c|c}
R_{2}^{2} & R_{3}^{2} & R_{4}^{2}\\\hline \end{array}\\
\begin{array}{c|c|c}
R_{2}^{2}=R_{3}^{2} & R_{3}^{2}=R_{4}^{2} & R_{2}^{2}=R_{4}^{2}\\\hline \end{array}\\
R_{2}^{2}=R_{3}^{2}=R_{4}^{2}
\end{array}$ & $\begin{array}{c}
\begin{array}{c|c|c}
\nicefrac{1}{2} & \nicefrac{1}{2} & \nicefrac{1}{3}\\\hline \end{array}\\
\begin{array}{c|c|c}
\nicefrac{1}{6} & \nicefrac{1}{3} & \nicefrac{1}{3}\\\hline \end{array}\\
\nicefrac{1}{6}
\end{array}$ & $\begin{array}{c}
\begin{array}{c|c|c}
\nicefrac{2}{3} & \nicefrac{2}{3} & \nicefrac{1}{2}\\\hline \end{array}\\
\begin{array}{c|c|c}
\nicefrac{1}{3} & \nicefrac{1}{3} & \nicefrac{1}{3}\\\hline \end{array}\\
\nicefrac{1}{6}
\end{array}$ & $\begin{array}{c}
\begin{array}{c|c|c}
\nicefrac{1}{2} & \nicefrac{1}{4} & \nicefrac{1}{4}\\\hline \end{array}\\
\begin{array}{c|c|c}
\nicefrac{1}{4} & 0 & 0\\\hline \end{array}\\
0
\end{array}$ & $\begin{array}{c}
\begin{array}{c|c|c}
\nicefrac{3}{5} & \nicefrac{2}{5} & \nicefrac{2}{5}\\\hline \end{array}\\
\begin{array}{c|c|c}
\nicefrac{1}{5} & \nicefrac{1}{5} & \nicefrac{2}{5}\\\hline \end{array}\\
\nicefrac{1}{5}
\end{array}$ & $\begin{array}{c}
\begin{array}{c|c|c}
\nicefrac{3}{5} & \nicefrac{2}{5} & \nicefrac{3}{5}\\\hline \end{array}\\
\begin{array}{c|c|c}
\nicefrac{1}{5} & \nicefrac{1}{5} & \nicefrac{2}{5}\\\hline \end{array}\\
\nicefrac{1}{5}
\end{array}$\tabularnewline
\hline 
$\begin{array}{c}
\begin{array}{c|c|c}
R_{3}^{3} & R_{4}^{3} & R_{1}^{3}\\\hline \end{array}\\
\begin{array}{c|c|c}
R_{3}^{3}=R_{4}^{3} & R_{4}^{3}=R_{1}^{3} & R_{3}^{3}=R_{1}^{3}\\\hline \end{array}\\
R_{3}^{3}=R_{4}^{3}=R_{1}^{3}
\end{array}$ & $\begin{array}{c}
\begin{array}{c|c|c}
\nicefrac{1}{2} & \nicefrac{1}{2} & \nicefrac{1}{3}\\\hline \end{array}\\
\begin{array}{c|c|c}
\nicefrac{1}{3} & \nicefrac{1}{3} & \nicefrac{1}{3}\\\hline \end{array}\\
\nicefrac{1}{6}
\end{array}$ & $\begin{array}{c}
\begin{array}{c|c|c}
\nicefrac{2}{3} & \nicefrac{1}{2} & \nicefrac{2}{3}\\\hline \end{array}\\
\begin{array}{c|c|c}
\nicefrac{1}{3} & \nicefrac{1}{3} & \nicefrac{1}{3}\\\hline \end{array}\\
\nicefrac{1}{6}
\end{array}$ & $\begin{array}{c}
\begin{array}{c|c|c}
\nicefrac{1}{2} & \nicefrac{1}{4} & \nicefrac{1}{2}\\\hline \end{array}\\
\begin{array}{c|c|c}
0 & \nicefrac{1}{4} & \nicefrac{1}{4}\\\hline \end{array}\\
0
\end{array}$ & $\begin{array}{c}
\begin{array}{c|c|c}
\nicefrac{2}{5} & \nicefrac{2}{5} & \nicefrac{3}{5}\\\hline \end{array}\\
\begin{array}{c|c|c}
0 & \nicefrac{1}{5} & \nicefrac{1}{5}\\\hline \end{array}\\
0
\end{array}$ & $\begin{array}{c}
\begin{array}{c|c|c}
\nicefrac{2}{5} & \nicefrac{3}{5} & \nicefrac{3}{5}\\\hline \end{array}\\
\begin{array}{c|c|c}
\nicefrac{1}{5} & \nicefrac{1}{5} & \nicefrac{1}{5}\\\hline \end{array}\\
0
\end{array}$\tabularnewline
\hline 
$\begin{array}{c}
\begin{array}{c|c|c}
R_{4}^{4} & R_{1}^{4} & R_{2}^{4}\\\hline \end{array}\\
\begin{array}{c|c|c}
R_{4}^{4}=R_{1}^{4} & R_{1}^{4}=R_{2}^{4} & R_{4}^{4}=R_{2}^{4}\\\hline \end{array}\\
R_{4}^{4}=R_{1}^{4}=R_{2}^{4}
\end{array}$ & $\begin{array}{c}
\begin{array}{c|c|c}
\nicefrac{1}{2} & \nicefrac{1}{2} & \nicefrac{1}{3}\\\hline \end{array}\\
\begin{array}{c|c|c}
\nicefrac{1}{3} & \nicefrac{1}{3} & \nicefrac{1}{3}\\\hline \end{array}\\
\nicefrac{1}{6}
\end{array}$ & $\begin{array}{c}
\begin{array}{c|c|c}
\nicefrac{1}{2} & \nicefrac{2}{3} & \nicefrac{2}{3}\\\hline \end{array}\\
\begin{array}{c|c|c}
\nicefrac{1}{3} & \nicefrac{1}{3} & \nicefrac{1}{3}\\\hline \end{array}\\
\nicefrac{1}{6}
\end{array}$ & $\begin{array}{c}
\begin{array}{c|c|c}
\nicefrac{3}{4} & \nicefrac{1}{4} & \nicefrac{1}{2}\\\hline \end{array}\\
\begin{array}{c|c|c}
\nicefrac{1}{4} & \nicefrac{1}{4} & \nicefrac{1}{2}\\\hline \end{array}\\
\nicefrac{1}{4}
\end{array}$ & $\begin{array}{c}
\begin{array}{c|c|c}
\nicefrac{3}{5} & \nicefrac{2}{5} & \nicefrac{3}{5}\\\hline \end{array}\\
\begin{array}{c|c|c}
\nicefrac{1}{5} & \nicefrac{1}{5} & \nicefrac{2}{5}\\\hline \end{array}\\
\nicefrac{1}{5}
\end{array}$ & $\begin{array}{c}
\begin{array}{c|c|c}
\nicefrac{3}{5} & \nicefrac{3}{5} & \nicefrac{3}{5}\\\hline \end{array}\\
\begin{array}{c|c|c}
\nicefrac{2}{5} & \nicefrac{1}{5} & \nicefrac{2}{5}\\\hline \end{array}\\
\nicefrac{1}{5}
\end{array}$\tabularnewline
\hline 
\end{tabular}

\end{adjustbox}

\caption{\label{tab:Distribution-of-the_hypercyclic-1-1}Distributions in the
selection of the disturbed hypercyclic systems used. The format is
the same as in Table \ref{tab:Distributions-of-the-1}.}
\end{table}

\FloatBarrier

\section*{Appendix B}

\setcounter{table}{0}
\renewcommand{\thetable}{B\arabic{table}}

Here, we present the distributions of the random variables in each
context of the systems $\mathcal{W}$ and $\mathcal{\mathcal{GHZ}}$
(Table \ref{tab:Distribution-of-variables}). Because our claim that
the $\mathcal{W}$ is noncontextual at level 2 seems to contradict
the claim made in Ref. \citep{Turaetal2014}, we present in Table
\ref{tab:Distribution-of-a} the distribution of a reduced coupling
of $\mathcal{W}$ from which one can compute the probabilities shown
in Table \ref{tab:Distribution-of-variables}. For instance, in Table
\ref{tab:Distribution-of-variables},
\[
\Prob\left[A_{1}^{2}=B_{2}^{2}=1\right]=\frac{1}{6}.
\]
By definition of the reduced coupling, we should have 
\[
\Prob\left[A_{1}^{2}=B_{2}^{2}=1\right]=\Prob\left[S_{1}=T_{2}=1\right].
\]
From Table \ref{tab:Distribution-of-a} we find 
\[
\Prob\left[S_{1}=T_{2}=1\right]=\Prob\left[\begin{array}{c}
S_{1}\\
\overline{S}_{2}\\
\overline{S}_{3}\\
T_{1}\\
T_{2}\\
T_{3}
\end{array}\right]+\Prob\left[\begin{array}{c}
S_{1}\\
\overline{S}_{2}\\
\overline{S}_{3}\\
\overline{T}_{1}\\
T_{2}\\
T_{3}
\end{array}\right]=\frac{1}{8}+\frac{1}{24}=\frac{1}{6}.
\]

\begin{table}
\begin{adjustbox}{max width = \textwidth}\arraycolsep=2pt\tabcolsep=5pt

\begin{tabular}{|c|c|c|}
\cline{2-3}
\multicolumn{1}{c|}{} & $\mathcal{W}$ & $\mathcal{\mathcal{GHZ}}$\tabularnewline
\hline 
$\begin{array}{c}
\begin{array}{c|c|c}
A_{1}^{1} & A_{2}^{1} & A_{3}^{1}\\\hline \end{array}\\
\begin{array}{c|c|c}
A_{1}^{1}=A_{2}^{1} & A_{2}^{1}=A_{3}^{1} & A_{1}^{1}=A_{3}^{1}\\\hline \end{array}\\
A_{1}^{1}=A_{2}^{1}=A_{3}^{1}
\end{array}$ & $\begin{array}{c}
\begin{array}{c|c|c}
\nicefrac{1}{3} & \nicefrac{1}{3} & \nicefrac{1}{3}\\\hline \end{array}\\
\begin{array}{c|c|c}
0 & 0 & 0\\\hline \end{array}\\
0
\end{array}$ & $\begin{array}{c}
\begin{array}{c|c|c}
\nicefrac{1}{2} & \nicefrac{1}{2} & \nicefrac{1}{2}\\\hline \end{array}\\
\begin{array}{c|c|c}
\nicefrac{1}{4} & \nicefrac{1}{4} & \nicefrac{1}{4}\\\hline \end{array}\\
\nicefrac{1}{4}
\end{array}$\tabularnewline
\hline 
$\begin{array}{c}
\begin{array}{c|c|c}
A_{1}^{2} & B_{2}^{2} & B_{3}^{2}\\\hline \end{array}\\
\begin{array}{c|c|c}
A_{1}^{2}=B_{2}^{2} & B_{2}^{2}=B_{3}^{2} & A_{1}^{2}=B_{3}^{2}\\\hline \end{array}\\
A_{1}^{2}=B_{2}^{2}=B_{3}^{2}
\end{array}$ & $\begin{array}{c}
\begin{array}{c|c|c}
\nicefrac{1}{3} & \nicefrac{1}{2} & \nicefrac{1}{2}\\\hline \end{array}\\
\begin{array}{c|c|c}
\nicefrac{1}{6} & \nicefrac{5}{12} & \nicefrac{1}{6}\\\hline \end{array}\\
\nicefrac{1}{12}
\end{array}$ & $\begin{array}{c}
\begin{array}{c|c|c}
\nicefrac{1}{2} & \nicefrac{1}{2} & \nicefrac{1}{2}\\\hline \end{array}\\
\begin{array}{c|c|c}
\nicefrac{1}{4} & \nicefrac{1}{4} & \nicefrac{1}{4}\\\hline \end{array}\\
0
\end{array}$\tabularnewline
\hline 
$\begin{array}{c}
\begin{array}{c|c|c}
B_{1}^{3} & A_{2}^{3} & B_{3}^{3}\\\hline \end{array}\\
\begin{array}{c|c|c}
B_{1}^{3}=A_{2}^{3} & A_{2}^{3}=B_{3}^{3} & B_{1}^{3}=B_{3}^{3}\\\hline \end{array}\\
B_{1}^{3}=A_{2}^{3}=B_{3}^{3}
\end{array}$ & $\begin{array}{c}
\begin{array}{c|c|c}
\nicefrac{1}{2} & \nicefrac{1}{3} & \nicefrac{1}{2}\\\hline \end{array}\\
\begin{array}{c|c|c}
\nicefrac{1}{6} & \nicefrac{1}{6} & \nicefrac{5}{12}\\\hline \end{array}\\
\nicefrac{1}{12}
\end{array}$ & $\begin{array}{c}
\begin{array}{c|c|c}
\nicefrac{1}{2} & \nicefrac{1}{2} & \nicefrac{1}{2}\\\hline \end{array}\\
\begin{array}{c|c|c}
\nicefrac{1}{4} & \nicefrac{1}{4} & \nicefrac{1}{4}\\\hline \end{array}\\
0
\end{array}$\tabularnewline
\hline 
$\begin{array}{c}
\begin{array}{c|c|c}
B_{1}^{4} & B_{2}^{4} & A_{3}^{4}\\\hline \end{array}\\
\begin{array}{c|c|c}
B_{1}^{4}=B_{2}^{4} & B_{2}^{4}=A_{3}^{4} & B_{1}^{4}=A_{3}^{4}\\\hline \end{array}\\
B_{1}^{4}=B_{2}^{4}=A_{3}^{4}
\end{array}$ & $\begin{array}{c}
\begin{array}{c|c|c}
\nicefrac{1}{2} & \nicefrac{1}{2} & \nicefrac{1}{3}\\\hline \end{array}\\
\begin{array}{c|c|c}
\nicefrac{5}{12} & \nicefrac{1}{6} & \nicefrac{1}{6}\\\hline \end{array}\\
\nicefrac{1}{12}
\end{array}$ & $\begin{array}{c}
\begin{array}{c|c|c}
\nicefrac{1}{2} & \nicefrac{1}{2} & \nicefrac{1}{2}\\\hline \end{array}\\
\begin{array}{c|c|c}
\nicefrac{1}{4} & \nicefrac{1}{4} & \nicefrac{1}{4}\\\hline \end{array}\\
0
\end{array}$\tabularnewline
\hline 
$\begin{array}{c}
\begin{array}{c|c|c}
A_{1}^{5} & A_{2}^{5} & B_{3}^{5}\\\hline \end{array}\\
\begin{array}{c|c|c}
A_{1}^{5}=A_{2}^{5} & A_{2}^{5}=B_{3}^{5} & A_{1}^{5}=B_{3}^{5}\\\hline \end{array}\\
A_{1}^{5}=A_{2}^{5}=B_{3}^{5}
\end{array}$ & $\begin{array}{c}
\begin{array}{c|c|c}
\nicefrac{1}{3} & \nicefrac{1}{3} & \nicefrac{1}{2}\\\hline \end{array}\\
\begin{array}{c|c|c}
0 & \nicefrac{1}{6} & \nicefrac{1}{6}\\\hline \end{array}\\
0
\end{array}$ & \tabularnewline
\hline 
$\begin{array}{c}
\begin{array}{c|c|c}
A_{1}^{6} & B_{2}^{6} & A_{3}^{6}\\\hline \end{array}\\
\begin{array}{c|c|c}
A_{1}^{6}=B_{2}^{6} & B_{2}^{6}=A_{3}^{6} & A_{1}^{6}=A_{3}^{6}\\\hline \end{array}\\
A_{1}^{6}=B_{2}^{6}=A_{3}^{6}
\end{array}$ & $\begin{array}{c}
\begin{array}{c|c|c}
\nicefrac{1}{3} & \nicefrac{1}{2} & \nicefrac{1}{3}\\\hline \end{array}\\
\begin{array}{c|c|c}
\nicefrac{1}{6} & \nicefrac{1}{6} & 0\\\hline \end{array}\\
0
\end{array}$ & \tabularnewline
\hline 
$\begin{array}{c}
\begin{array}{c|c|c}
B_{1}^{7} & A_{2}^{7} & A_{3}^{7}\\\hline \end{array}\\
\begin{array}{c|c|c}
B_{1}^{7}=A_{2}^{7} & A_{2}^{7}=A_{3}^{7} & B_{1}^{7}=A_{3}^{7}\\\hline \end{array}\\
B_{1}^{7}=A_{2}^{7}=A_{3}^{7}
\end{array}$ & $\begin{array}{c}
\begin{array}{c|c|c}
\nicefrac{1}{2} & \nicefrac{1}{3} & \nicefrac{1}{3}\\\hline \end{array}\\
\begin{array}{c|c|c}
\nicefrac{1}{6} & 0 & \nicefrac{1}{6}\\\hline \end{array}\\
0
\end{array}$ & \tabularnewline
\hline 
$\begin{array}{c}
\begin{array}{c|c|c}
B_{1}^{8} & B_{2}^{8} & B_{3}^{8}\\\hline \end{array}\\
\begin{array}{c|c|c}
B_{1}^{8}=B_{2}^{8} & B_{2}^{8}=B_{3}^{8} & B_{1}^{8}=B_{3}^{8}\\\hline \end{array}\\
B_{1}^{8}=B_{2}^{8}=B_{3}^{8}
\end{array}$ & $\begin{array}{c}
\begin{array}{c|c|c}
\nicefrac{1}{2} & \nicefrac{1}{2} & \nicefrac{1}{2}\\\hline \end{array}\\
\begin{array}{c|c|c}
\nicefrac{5}{12} & \nicefrac{5}{12} & \nicefrac{5}{12}\\\hline \end{array}\\
\nicefrac{3}{8}
\end{array}$ & \tabularnewline
\hline 
\end{tabular}

\end{adjustbox}

\caption{\label{tab:Distribution-of-variables}Distribution of variables in
the $\mathcal{W}$ and $\mathcal{\mathcal{GHZ}}$systems. The format
is the same as in Table \ref{tab:Distributions-of-the-1}}
\end{table}

\begin{table}
\begin{adjustbox}{max width = \textwidth}\arraycolsep=2pt\tabcolsep=5pt%
\begin{tabular}{|c|c|c|c|c|c|c|c|c|c|c|c|c|c|}
\hline 
\begin{tabular}{c}
value\tabularnewline
of\tabularnewline
coupling\tabularnewline
\end{tabular} & $\begin{array}{c}
S_{1}\\
\overline{S}_{2}\\
\overline{S}_{3}\\
T_{1}\\
T_{2}\\
T_{3}
\end{array}$ & $\begin{array}{c}
S_{1}\\
\overline{S}_{2}\\
\overline{S}_{3}\\
\overline{T}_{1}\\
T_{2}\\
T_{3}
\end{array}$ & $\begin{array}{c}
S_{1}\\
\overline{S}_{2}\\
\overline{S}_{3}\\
T_{1}\\
\overline{T}_{2}\\
\overline{T}_{3}
\end{array}$ & $\begin{array}{c}
S_{1}\\
\overline{S}_{2}\\
\overline{S}_{3}\\
\overline{T}_{1}\\
\overline{T}_{2}\\
\overline{T}_{3}
\end{array}$ & $\begin{array}{c}
\overline{S}_{1}\\
S_{2}\\
\overline{S}_{3}\\
T_{1}\\
T_{2}\\
T_{3}
\end{array}$ & $\begin{array}{c}
\overline{S}_{1}\\
S_{2}\\
\overline{S}_{3}\\
\overline{T}_{1}\\
\overline{T}_{2}\\
\overline{T}_{3}
\end{array}$ & $\begin{array}{c}
\overline{S}_{1}\\
\overline{S}_{2}\\
S_{3}\\
T_{1}\\
T_{2}\\
T_{3}
\end{array}$ & $\begin{array}{c}
\overline{S}_{1}\\
\overline{S}_{2}\\
S_{3}\\
T_{1}\\
T_{2}\\
\overline{T}_{3}
\end{array}$ & $\begin{array}{c}
\overline{S}_{1}\\
\overline{S}_{2}\\
S_{3}\\
\overline{T}_{1}\\
T_{2}\\
\overline{T}_{3}
\end{array}$ & $\begin{array}{c}
\overline{S}_{1}\\
\overline{S}_{2}\\
S_{3}\\
T_{1}\\
\overline{T}_{2}\\
T_{3}
\end{array}$ & $\begin{array}{c}
\overline{S}_{1}\\
\overline{S}_{2}\\
S_{3}\\
\overline{T}_{1}\\
\overline{T}_{2}\\
T_{3}
\end{array}$ & $\begin{array}{c}
\overline{S}_{1}\\
\overline{S}_{2}\\
S_{3}\\
\overline{T}_{1}\\
\overline{T}_{2}\\
\overline{T}_{3}
\end{array}$ & $\ldots$\tabularnewline
\hline 
$\Prob$ & $\nicefrac{1}{8}$ & $\nicefrac{1}{24}$ & $\nicefrac{1}{24}$ & $\nicefrac{1}{8}$ & $\nicefrac{1}{6}$ & $\nicefrac{1}{6}$ & $\nicefrac{1}{12}$ & $\nicefrac{1}{24}$ & $\nicefrac{1}{24}$ & $\nicefrac{1}{24}$ & $\nicefrac{1}{24}$ & $\nicefrac{1}{12}$ & 0\tabularnewline
\hline 
\end{tabular}

\end{adjustbox}

\caption{\label{tab:Distribution-of-a}Distribution of a reduced coupling $\left(S_{1},S_{2},S_{3},T_{1},T_{2},T_{3}\right)$
of system $\mathcal{W}$ (with $S$ and $T$ being distributional
copies of $A$ and $B,$ respectively). The events whose probabilities
are shown in the bottom row are values of the reduced coupling $\left[S_{1}=\pm1,S_{2}=\pm1,S_{3}=\pm1,T_{1}=\pm1,T_{2}=\pm1,T_{3}=\pm1\right],$
shown in the following way: $S_{i}$ indicates $S_{i}=1$ and $\overline{S}_{i}$
indicates $S_{i}=-1$; analogously for $T_{j}$ and $\overline{T}_{j}$.
All values of the reduced coupling not explicitly shown have probability
zero.}
\end{table}

\FloatBarrier

\paragraph*{Acknowledgements:}

We are indebted to Pawe{\l} Kurzy{\'n}ski for critically reading
the first draft of this paper and making valuable suggestions.

\paragraph*{Author Contributions:}

The initial idea: END. Numerical analysis and software: VHC. The authors
contributed equally to all other aspects of the work. 

\paragraph*{Conflicts of Interest:}

The authors declare no conflict of interest.

\end{document}